\documentclass[iop]{emulateapj}
\usepackage{graphicx,amsmath,amsfonts,amssymb,subfigure}

\newcommand{\kms}{\,~km~s$^{-1}$}

\def\rsun{\ifmmode {\rm R}_{\mathord\odot}\else $R_{\mathord\odot}$\fi}
\def\msun{\ifmmode {\rm M}_{\mathord\odot}\else $M_{\mathord\odot}$\fi}
\def\lsun{\ifmmode {\rm L}_{\mathord\odot}\else $L_{\mathord\odot}$\fi}

\newcommand{\nhtd}{\mbox{\rm NH$_2$D(1$_{1,1}$-1$_{0,1}$)}}
\newcommand{\nthp}{\mbox{\rm N$_2$H$^+$(1-0)}}
\newcommand{\hcn}{\mbox{\rm HCN(1-0)}}
\newcommand{\hcop}{\mbox{\rm  HCO$^+$(1-0)}}
\newcommand{\cs}{\mbox{\rm CS(2-1)}}

\slugcomment{\it Accepted for publication in ApJ Dec 3, 2016}

\begin{document}

\title{Kinematics of a young low-mass star forming core: Understanding the evolutionary state of the First Core Candidate L1451-\MakeLowercase{mm}}

\author{Mar\'ia Jos\'e Maureira, H\'ector G. \ Arce}
\affil{Astronomy Department, Yale University, New Haven, CT~06511, USA}
\email{mariajose.maureira@yale.edu, hector.arce@yale.edu}

\author{Michael M. Dunham}
\affil{Department of Physics, State University of New York at Fredonia, Fredonia, NY 14063, USA}

\author{Jaime E. Pineda}
\affil{Max-Planck Institute for Extraterrestrial Physics, Giessenbachstrasse 1, 85748 Garching, Germany}

\author{Manuel Fern\'andez-L\'opez}
\affil{Instituto Argentino de Radioastronom\'ia, CCT-La Plata (CONICET), C.C.5, 1894, Villa Elisa, Argentina}

\author{Xuepeng Chen}
\affil{Purple Mountain Observatory, Chinese Academy of Sciences, 2 West Beijing Road, Nanjing 210008, China}

\and

\author{Diego Mardones}
\affil{Departamento de Astronom\'ia, Universidad de Chile, Casilla 36-D, Santiago, Chile}

\begin{abstract}

We use 3mm multi-line and continuum CARMA observations towards the first hydrostatic core (FHSC) candidate L1451-mm to characterize the envelope kinematics at 1000 AU scales and investigate its evolutionary state. We detect evidence of infall and rotation in the \nhtd, \nthp\ and \hcn\ molecular lines. We compare the position velocity diagram of the \nhtd\ line with a simple kinematic model and find that it is consistent with an envelope that is both infalling and rotating while conserving angular momentum around a central mass of about $0.06$ M$_{\odot}$. The \nthp\ LTE mass of the envelope along with the inferred infall velocity leads to a mass infall rate of approximately $6\times10^{-6}$ M$_{\odot}$ yr$^{-1}$, implying a young age of $10^4$ years for this FHSC candidate. Assuming that the accretion onto the central object is the same as the infall rate we obtain that the minimum source size is 1.5-5 AU —consistent with the size expected for a first core. We do not see any evidence of outflow motions or signs of outflow-envelope interaction at scales $\ga$ 2000 AU. This is consistent with previous observations that revealed a very compact outflow ($\la$ 500 AU). We conclude that L1451-mm is indeed at a very early stage of evolution, either a first core or an extremely young Class 0 protostar. Our results provide strong evidence that L1451-mm is the best candidate for being a bonafide first core.

\end{abstract}


\section{Introduction}
\label{sec_intro}

The formation of a low mass star begins when a dense core (\citealt{1989BensonSurvey,2002CaselliDense}) becomes unstable under its own gravity and starts to collapse (\citealt{1969LarsonNumerical,1987ShuStar}). When the dense inner parts become opaque to radiation the thermal pressure is able to balance the gravity and a central structure, of a few AU in size and $0.01-0.1$ M$_{\odot}$ in mass, reaches quasi-hydrostatic equilibrium. This structure is known as the first hydrostatic core (FHSC) or first core (\citealt{1969LarsonNumerical,1998Masunaga,2006SaigoEvolution,2013TomidaRadiation,2014BateCollapse,2015TomidaRadiation}).  The first core continues accreting material, increasing both its central density and temperature for the next $10^{3}-10^{4}$ years. When the central temperature reaches approximately $2000$ K, the molecular hydrogen dissociates in an endothermic reaction, triggering a second collapse in a small region (about 0.1 AU) inside the first core. This second collapse leads to the formation of a second hydrostatic core, namely a protostar. During its first years the protostar has a radius of a few $R_{\odot}$ and a mass of only few $0.001-0.01$ M$_{\odot}$. After the protostar is born, it grows significantly in mass by accreting material from the first core and later from the remaining core or envelope (\citealt{2010InutsukaEmergence,2014BateCollapse}). Material will be transported from the envelope into the protostar through a rotationally supported disk (\citealt{2009JorgensenPROSAC,2015TobinSub,2015VorobyovVariable}) while a fraction of the infalling mass will be ejected back to the envelope through bipolar winds or outflows (\citealt{2014OffnerInvestigations,2015PlunkettEpisodic}). After $\sim0.5$ Myrs, most of the envelope has been accreted or ejected and the protostar becomes a Pre-Main Sequence star \citep{2009EvansSpitzer,2014DunhamEvolution}.\\

Observational evidence regarding these early stages of low mass star formation come from the study of collapsing (prestellar) cores and young (Class 0) protostars (reviews in \citealt{2000AndreFrom, 2007diFrancescoObservational,2007WardThompsonObservational,2014DunhamEvolution}). However, many of the details mentioned above come only from simulations (\citealt{2011MatsumotoProtostellar,2013JoosInfluence, 2013TomidaRadiation, 2014BateCollapse, 2014LiEarliest,2015TsukamotoEffects,2015TomidaRadiation} and references therein). The FHSC although predicted more than 40 years ago \citep{1969LarsonNumerical}, has never been observationally confirmed. This early stage is important as the formation and growth of both the circumstellar disk and outflows are predicted to occur between the first core phase and the first $10^{4}$ years into the protostellar phase. For instance, simulations predict the formation of an early outflow, low in velocity ($\lesssim10$ \kms) and poorly collimated, launched by the first core rather than the protostar \citep{2011MatsumotoProtostellar, 2013TomidaRadiation, 2014MachidaProtostellar, 2014BateCollapse,2015TomidaRadiation}. Once the protostar is formed, a second outflow, highly collimated and with high velocities ($\gtrsim10$ \kms), is predicted to be launched by the protostar \citep{2013TomidaRadiation, 2014MachidaProtostellar, 2014BateCollapse,2015TomidaRadiation}.\\

In numerical simulations, the formation and properties of outflow and rotationally supported disks vary with the physics that is included in the model (e.g., radiative transfer, ideal/non ideal MHD, etc), the  initial conditions (rotation, density distribution, symmetry, magnetic field and turbulence strength, etc.) and resolution. The development of a disk might start right after protostar formation \citep{2010InutsukaEmergence, 2011MachidaOrigin,2015TsukamotoEffects}, or a rotationally supported structure corresponding to the first core, or surrounding it, might exist before protostar formation \citep{2010InutsukaEmergence, 2011MachidaOrigin,2015TsukamotoEffects,2012JoosProtostellar, 2013TsukamotoFormation,2016MassonAmbipolar}. Observations that can test these predictions are needed in order to better understand the formation of protostars and their evolution.\\

Among the observational classes of young objects that include Class 0 sources and very low luminosity objects (VeLLOs), FHSC candidates are an ideal target for studying the earliest stages of low mass star formation; by studying them we can test the existence of first cores as well as learn about the formation and early evolution of protostars, disks and outflows. Based on SED properties, past studies have reported about nine FHSC candidates (e.g. \citealt{2006BellocheEvolutionary,2010ChenL1448,2010EnochCandidate,2011DunhamDetection,2011PinedaEnigmatic,2012ChenSubmillimeter,2012SchneeHow,2012PezzutoHerschel, 2013HuangProbing, 2013MurilloDisentangling}). These SEDs indicate that these sources are lower in luminosity and more deeply embedded than typical young Class 0 protostars, including VeLLOs, and are also consistent with SEDs from simulations of first cores. Furthermore, some of the first core candidates drive small outflows, with properties that are in agreement with simulations of FHSC. Despite this evidence, probing the true nature of these candidates (whether they are bona fide first cores or very young low-luminosity protostars) has proven to be difficult.\\

Observational studies of the kinematics of the gas surrounding young protostars are useful to further characterize the properties of these objects. Such studies have been done for Class 0 sources probing the velocity and linewidth distribution as well as infall and rotational motions from tens to thousands of AU scales (\citealt{1997OhashiInterferometric,2011TobinComplex,2007ChenOvro,2013MurilloKeplerian,2013YenUnveiling,2014LindbergAlma,2014MaretFirst,2014OyaSubstellar,2015YenNo}). Comparing the kinematics of envelopes around first core candidates with those of known (young) Class 0 sources may help understand the nature of these interesting sources. For instance, \cite{2013TsitaliDynamical} studied the first core candidate Cham-MMS1 and found envelope infall velocities of $\sim$0.2 \kms,  an average gradient of 3 km s$^{-1}$ pc$^{-1}$ at 1000 AU scales and no signs of a fast, large scale outflow. They concluded that the envelope kinematic properties of this source are consistent with either a first core or a very young Class 0 protostar (see also \citealt{2014VaisalaHigh}).\\

In this study, we present CARMA (Combined Array for Research in Millimeter-wave) molecular line observations of the first core candidate L1451-mm, using the dense gas tracers: \nhtd, \nthp, \hcn, \hcop\ and \cs. Our goal is to characterize the kinematics of the envelope at 1000 AU scales in order to compare with observations of Class 0 sources and thus, further constrain the evolutionary state of this candidate. 

L1451-mm is a dense core in the L1451 dark cloud located in the Perseus Molecular Cloud at a distance of $\sim230$ pc \citep{2008HirotaAstrometry}. Interferometric millimeter observations of 1" resolution show a 1.3 continuum point source at the center of the core, and no IR counterpart detected \citep{2011PinedaEnigmatic}. The source drives a low-velocity ($\sim2$ \kms) compact outflow with a dynamical time of just 2000 years and has an SED consistent with that expected for an FHSC. These properties were used by \cite{2011PinedaEnigmatic} to classify this source as a first core candidate. In addition, \cite{2011PinedaEnigmatic} inferred an upper limit for the internal luminosity of only 0.016 $L_{\odot}$, one of the lowest among the first cores candidates.\\

The paper is organized as follows: in Section 2, we describe the observations and data reduction; in Section 3, we present our general results; and in Section 4, we estimate masses and analyze the kinematics of L1451-mm, quantifying infall and rotational motions. In Section 5, we compare our results with those of studies of Class 0 sources and discuss the implications of our results regarding the evolutionary state of L1451-mm. 

\section{Observations and reduction}

Observations of L1451-mm for this study were conducted using the CARMA array between April and August 2012, using the two most compact configurations (D and E) which provided baselines from 8 to 150 m, and thus sensitivity to structures up to 50" ($\sim$12000 AU). A typical track included observations of a flux calibrator at the beginning, observation of a bandpass calibrator at the beginning or end of the track and several science source-phase calibrator cycles in the middle. These cycles were organized such that every observation of the science source (9.7 minutes each) was bracketed by a phase calibrator observation (3.9 minutes each). Uranus was selected for flux calibration, 0336+323 for phase calibration  and 3c84 or  0423-013 for bandpass calibration.  The total integration time for L1451-mm was 3.1 and 2.4 hours in D and E configuration, respectively. 

The correlator configuration consisted of seven 8 MHz windows with 319 channels ( $\Delta v \sim 0.085$ \kms) for spectral lines and two 500 MHz windows for continuum at $\sim$ 90 GHz. The seven narrow windows were centered in the following molecular transitions: \nhtd, \nthp, \hcn, \hcop, \cs, SiO(2-1), and C$^{34}$S(2-1). The range of system temperatures during the tracks was from 130 to 270 K with a variation within the same antenna of 20-50 K, consistent with changes due to elevation. \\

The software MIRIAD was used for both calibration and imaging.  We performed a standard calibration consisting of baseline determination, line length calibration,  flagging, passband calibration, and amplitude and phase calibration. For imaging, the task mosmem was used for the deconvolution and the channels were re-sampled to a resolution of 0.1\kms. Continuum emission and emission from all but the SiO(2-1), and C$^{34}$S(2-1) spectral lines were detected. 
We combined our single-pointing observations of L1451-mm (described above) with previous 7-pointing mosaic CARMA observations in the C, D and E configuration for \nthp, in D and C configurations for \hcop\ and 3mm continuum, and in D configuration for \nhtd, presented in \cite{2011PinedaEnigmatic}. Combining our CARMA D and E observations with the C-configuration observations added baselines of up to 350 m. The sizes of the final synthesized beams are about 5" to 6", which at the distance of the Perseus molecular clouds (230 pc) corresponds to about 1270 AU. Table~\ref{tb:l1451-mm_maps} lists the line frequencies, array configurations, beam sizes and RMS of the final L1451-mm data cubes.

\begin{deluxetable*}{lcclrc}
\tablecaption{L1451-\textrm{\normalfont mm} Maps parameters  \label{tb:l1451-mm_maps} }
\tablewidth{450pt}
\tablehead{
\colhead{Map} & \colhead{Rest Frequency}& \colhead{Array configurations}& \colhead{Synthesized beam} & \colhead{PA} & \colhead{RMS} \\
\colhead{}&\colhead{[GHz]}&\colhead{}&\colhead{}&\colhead{}&\colhead{[mJy beam$^{-1}$]}}
\startdata
NH$_2$D$(1_{1,1}-1_{0,1})$ & 85.926263 & D+E & 6".7 x 5".8&$+75.4^{\circ}$ & 75 \\ 
N$_2$H$^+(1-0)$&93.17340 &C+D+E& 5".9 x 5".0&$-83.9^{\circ}$ &  55 \\ 
HCN$(1-0)$ &88.631846&D+E& 6".6 x 5".5&$+78.6^{\circ}$  & 56 \\ 
HCO$^+(1-0)$&89.18853&C+D+E&  5".7 x 4".9&$-87.8^{\circ}$  & 86 \\ 
CS$(2-1)$ &97.98095&D+E& 6".1 x 5".1&$+70.2^{\circ}$   & 67 \\ 
3mm continuum &90.58956&C+D+E& 5".5 x 4".8&$+89.9^{\circ}$ & 0.6 \\ 
\enddata
\tablecomments{ The RMS for the molecular lines is measured using channels that are 0.1 \kms\ wide.}
\end{deluxetable*}

\section{Results}

\begin{figure*}
\includegraphics[width=1\textwidth]{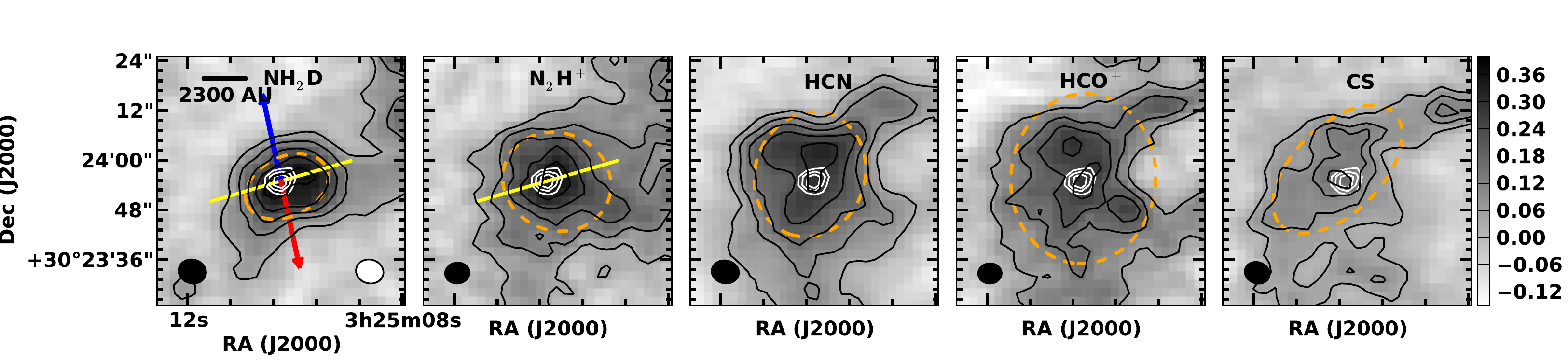}
\caption{L1451-mm integrated intensity maps. The integration velocity range is  [3.7,4.6] \kms\ for all maps, except for that of \cs\ which  is [3.5,3.9] \kms. In the case of lines with hyperfine structure, this range corresponds to integration that includes only the most intense hyperfine component. Black contours are drawn at 10, 30, 50, 70 and 90 \% of the maximum peak integrated intensity for \nhtd, \nthp, and \hcn, and at 20, 50, 70 and 90 \% of the maximum peak integrated intensity for \hcop\ and \cs.  The maximum values of the maps are listed in Table~\ref{tb:l1451-mmproperties}. The lowest contour corresponds to at least 2 sigma for all the maps. White contours show the 3mm continuum at 2, 3 and 4 sigma. The red and blue arrows show the directions of the red and blueshifted outflow lobes observed by \cite{2011PinedaEnigmatic}. The orange dashed line in each case is the ellipse corresponding to the FWHM obtained from a gaussian fit to the molecular line integrated intensity map (see Table~\ref{tb:l1451-mmproperties}). The yellow line shows the cut used for the position velocity maps in Figure~\ref{fig:L1451-mmposvel}. \label{fig:L1451-mmmom0}}
\end{figure*}

\begin{figure}
\includegraphics[width=0.48\textwidth]{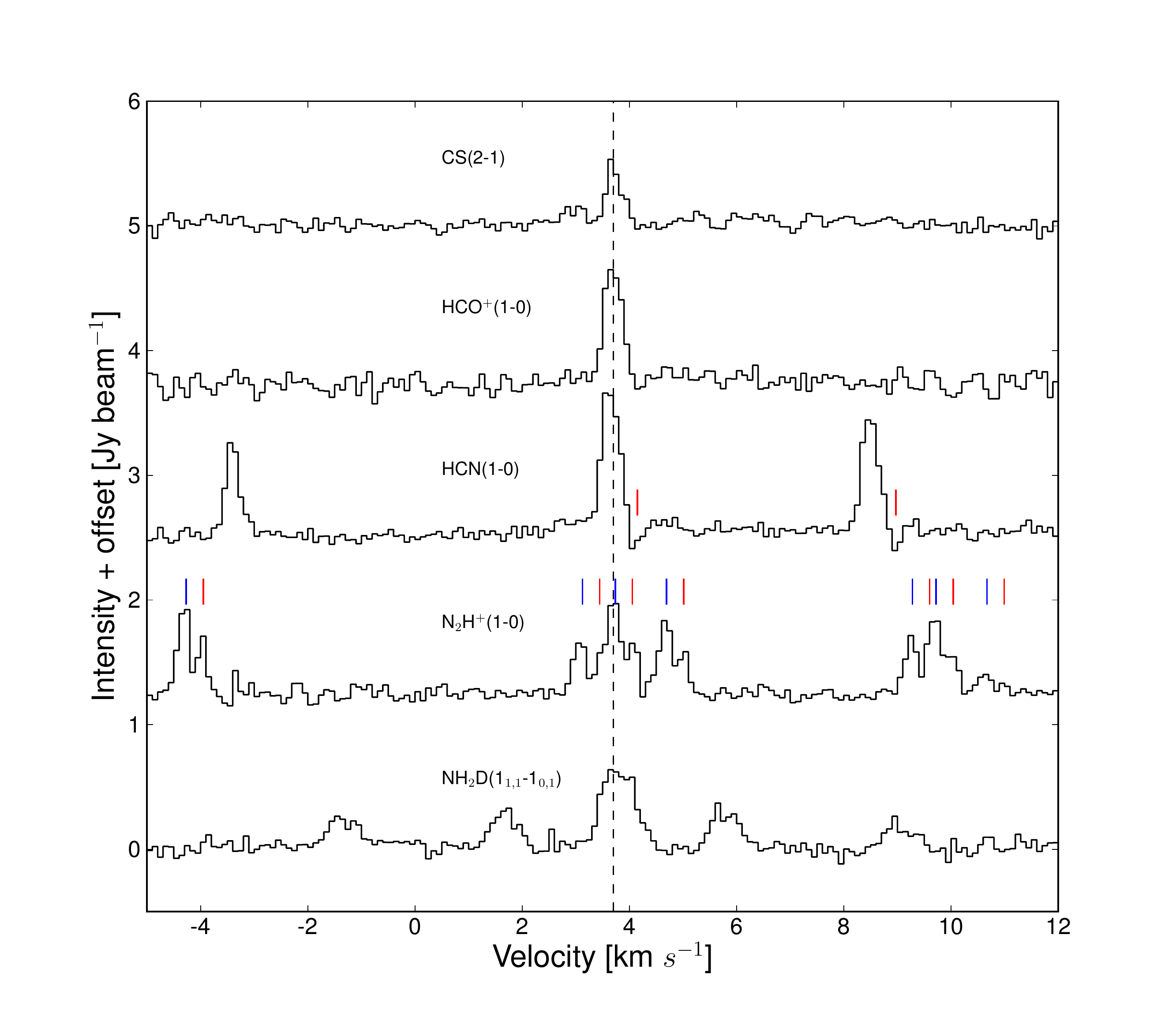}
\caption{L1451-mm molecular line spectra at the continuum peak position, averaged over a beam. The two components of the double peak profile for the \nthp\ are marked with red and blue lines, while the redshifted absorptions in \hcn\ are marked with red lines. The vertical line is at 3.9 \kms. This value indicates the envelope systemic velocity as given by a fit to the entire velocity map of the \nhtd\ and \nthp\ (Figure~\ref{fig:l1451-mmvel}). See Section 4.2.1 for more details on the fit to the velocity maps.\label{fig:L1451-mmspec}}
\end{figure}

Figure~\ref{fig:L1451-mmmom0} shows integrated intensity maps for all the molecules with detected emission. The 3mm continuum emission is also shown in white contours. We fit the integrated intensity maps, for all the molecules and the continuum with a 2D-Gaussian fit. We considered only pixels above 3$\sigma$ in the integrated intensity map and inside a square region of 30" at the center of the map (where the vast majority of the emission is concentrated). Table~\ref{tb:l1451-mmproperties} lists the size, position angle and aspect ratio of the molecular emission obtained from the fit. The dashed orange lines in Figure~\ref{fig:L1451-mmmom0} correspond to ellipses with the major and minor axis equal to the FWHM values obtained from the fit and oriented according to the position angle obtained from the fit. The size, visually or measured as the geometric mean of the FWHM of the deconvolved major and minor axes of the integrated intensity, increases from \nhtd\ to \hcop. The structure traced by the \cs\ is slightly more compact than the one traced by \hcop. The \nhtd\ and \nthp\ emission exhibit the two most compact structures of all the molecular line maps and have a semi-major axis within 15 degree and 34 degree of the direction perpendicular to the outflow, respectively. In all the molecular line maps the continuum emission lies within the contour representing 90\% of the peak emission of the integrated intensity map. Moreover, the separation between the continuum peak and the peak of the integrated intensity map is within 2-4" (less than the size of the synthesized beam). All molecular line maps show emission extending toward a particular direction. For example, the structures traced by the nitrogen-bearing molecules \nhtd\ and \nthp, have a tail extending toward the West while the structures traced by the carbon-bearing molecules \hcn, \hcop\ and \cs\ have a tail toward the North-West. Emission extending towards the South is common to all molecular line maps. \\

\begin{figure}
\includegraphics[width=0.48\textwidth]{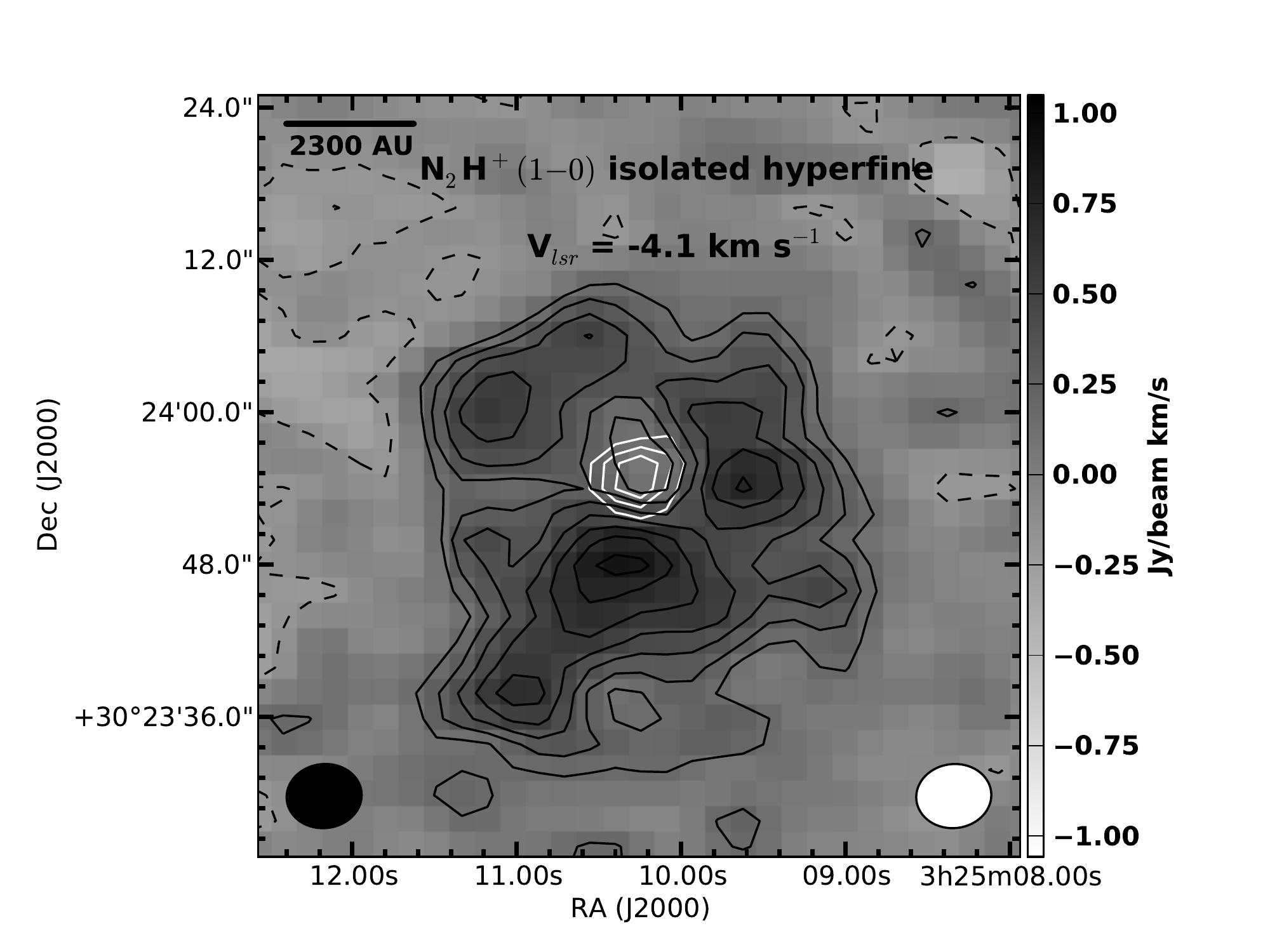}
\caption{ \nthp\ emission at the velocity channel ($v_{lsr}=4.1$ \kms) corresponding to the dip of the double peak profile in the isolated hyperfine spectrum (Figure~\ref{fig:L1451-mmspec}). Black contours follow the molecular line emission starting at 3$\sigma$ and increasing in steps of 2$\sigma$. White contours show the continuum emission at 2, 3 and 4$\sigma$. The \nthp\ synthesized beam is shown in black and the continuum synthesized beam is shown in white.\label{fig:L1451-mmhole}}
\end{figure}

Figure~\ref{fig:L1451-mmspec} shows spectra at the continuum position, averaged over a beam, for all the observed molecular lines. All spectra shown in Figure~\ref{fig:L1451-mmspec}, except that of \nthp, present a single peak structure. The \nthp\ line shows a double peak asymmetric structure with the blue peak (i.e., the peak at lower LSR velocities) brighter than the red peak, which is more clearly seen in the isolated hyperfine component at $v\sim-4$ \kms. The red and blue velocity components for each hyperfine component are marked with a blue and red line in Figure~\ref{fig:L1451-mmspec}, respectively. This type of profile is found in the presence of infall \citep{1997MardonesSearch}. This double peak structure in the \nthp\ spectrum is only seen at the position of the continuum. In the case of  \hcn, its spectrum presents a faint absorption feature at (small) redshifted velocities with respect to the central envelope velocity (see the red mark in Figure~\ref{fig:L1451-mmspec}). Similar to \nthp, the red absorption is only seen at the position of the continuum source.
Both absorptions are 4 to 5 channels wide (i.e., $0.5$ \kms). The absorption peak to the red of the hyperfine component at $v\sim3.7$ \kms\ has a S/N of 2.5, while the absorption to the red of the hyperfine component at $v\sim8.5$ \kms\ has a S/N of 3. This profile shape (or inverse P Cygni profile) is indicative of infall motions seen against continuum emission (e.g., \citealt{2001diFrancescoInfall,2012PinedaFirst}). Although the absorption in the inverse P Cygni profile is faint, it is unlikely that missing flux due to filtering of large-scale structure in our interferometric observations is responsible for the absorption feature since both this and the double peak profile in the \nthp\ spectrum neatly coincide with the position of the continuum source. Figure~\ref{fig:L1451-mmhole} shows the \nthp\ emission channel corresponding to the dip of the double peak profile, where fainter emission is seen at the position of the continuum white contours. \\

The \hcop\ and \cs\ lines peak  close to the position of the blue peak of \nthp. However, none of them show an asymmetry that could be conclusively associated with infall motions. 

\begin{deluxetable*}{llccc}
\tabletypesize{\scriptsize}
\tablewidth{450pt} 
\tablecaption{L1451-\textrm{\normalfont mm} Molecular line integrated intensity and continuum map properties \label{tb:l1451-mmproperties}  }
\tablehead{
\colhead{Map} & \colhead{Size$^{ a}$} & \colhead{PA$^{b}$} & \colhead{Aspect Ratio$^{c}$} & \colhead{Peak value$^{d}$ }  \\
\colhead{} & \colhead{["]}&\colhead{[$^{\circ}$]} &\colhead{} &\colhead{}}
\startdata
NH$_2$D$(1_{1,1}-1_{0,1})$ & 17.4 $\pm$ 0.2 & -61 $\pm$ 2 & 1.5 & 0.39 $\pm$ 0.02  \\
N$_2$H$^+(1-0)$&25.0 $\pm$ 0.4 & 70 $\pm$7 & 1.1 & 0.36 $\pm$ 0.02  \\
HCN$(1-0)$ & 28.4 $\pm$ 0.5 & -18 $\pm$ 9& 1.1       &0.34 $\pm$ 0.02   \\
HCO$^+(1-0)$& 37.5 $\pm$ 1.3&  -7 $\pm$ 8& 1.2    &0.30 $\pm$ 0.03 \\
CS$(2-1)$ & 28.5 $\pm$ 0.8& -46 $\pm$ 3 &    1.8 & 0.17 $\pm$ 0.01 \\
3mm continuum & 4.5 $\pm$ 0.9 & -44 $\pm$ 15 & 2.5  &0.003 $\pm$ 0.0004     \\
\enddata
\tablecomments{ The properties: size, PA and aspect ratio were obtained by fitting the integrated emission maps in Figure~\ref{fig:L1451-mmmom0} with a 2D Gaussian.}
\tablenotetext{a}{This value corresponds to the geometric mean of the deconvolved major FWHM axis and minor FWHM axis from the gaussian fit. }
\tablenotetext{b}{Deconvolved position angle (East of North) of the emission obtained from a 2D-Gaussian fit.}
\tablenotetext{c}{Ratio between the deconvolved major FWHM axis and the minor FWHM axis.}
\tablenotetext{d}{Peak of the integrated intensity map. The molecular line peak values are in units of Jy beam$^{-1}$ km s$^{-1}$, while the peak value for the continuum is in units of Jy beam$^{-1}$.}
\end{deluxetable*}

\section{Analysis}

\subsection{Line Fitting}

\begin{figure*}
\includegraphics[width=1\textwidth]{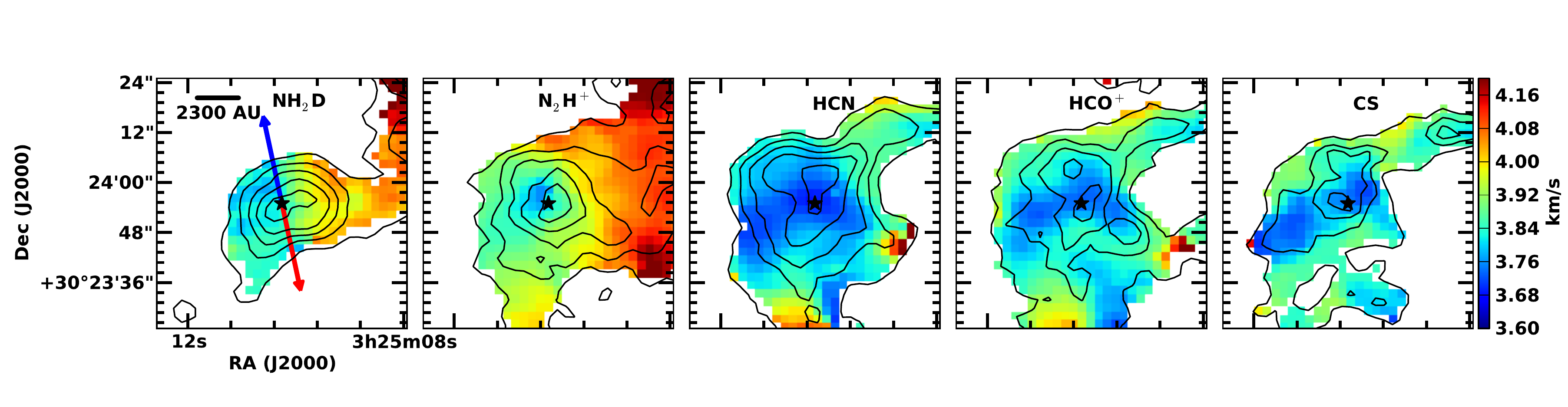}
\caption{L1451-mm central velocity maps. The maps show the central velocity obtained by fitting the spectrum with a gaussian profile at each position where the integrated intensity is greater than 3 sigma (see Sec. 3). The black contours are the same as in Figure ~\ref{fig:L1451-mmmom0}. The black star marks the position of the continuum point source. The red and blue arrows show the directions of the red and blueshifted outflow lobes observed by \cite{2011PinedaEnigmatic}.
\label{fig:l1451-mmvel}}
\end{figure*}

\begin{figure*}
\includegraphics[width=1\textwidth]{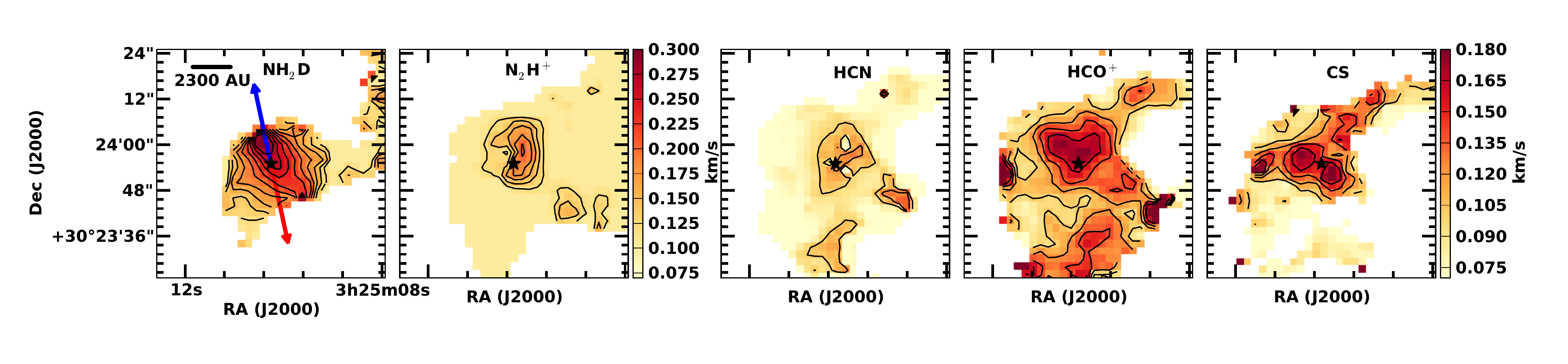}
\caption{L1451-mm linewidths maps. The maps show the linewidths $\sigma$ obtained by fitting the spectrum with a Gaussian profile at each position where the integrated intensity is greater than 3 sigma (see Sec. 3). The black contours starts at 0.1 \kms and increases in steps of 0.02 \kms. The black star marks the position of the continuum point source. The red and blue arrows show the directions of the red and blueshifted outflow lobes observed by \cite{2011PinedaEnigmatic}.
\label{fig:l1451-mmsigma}}
\end{figure*}

In order to analyze the velocities and linewidths we fit the molecular line spectra with a Gaussian profile. For spectra with hyperfine structure,  the intensity as a function of velocity is fit by:\\

\begin{equation}
I_v=I_0\left[\frac{1}{e^{T_0/T_{ex}}-1}-\frac{1}{e^{T_0/T_{cmb}}-1}\right](1-e^{-\tau_v})
\label{eq:gaussianfit}
\end{equation}

where $T_{ex}$ is the excitation temperature, $T_0=h\nu/k$, $I_0=2h\nu^3/c^2$ and $\nu=(1-v/c)\nu_{0}$ with $\nu_{0}$ the rest frequency of the main hyperfine line. The opacity term $\tau_{v}$ corresponds to:
\begin{equation}
\tau_{v}=\tau_{tot}\cdot\sum_{i}\tau_{i}e^{-\frac{(v-(v_{c}+v_i))^2}{2\sigma_v^2}}
\label{eq:gaussianfit2}  
\end{equation}
where the sum is over the hyperfine line components. Here, $v_{c}$ is the observed central velocity, $\sigma_v$ is the linewidth and  $\tau_{tot}$ is the total opacity. The parameters $T_{ex}$, $v_c$, $\sigma_v$ and $\tau_{tot}$ and their errors are directly obtained from a non-linear least square fitting routine that uses the Python package lmfit. The parameter $\tau_i$ corresponds to the fraction of the total opacity of the $i$-th hyperfine line component with $\sum_i \tau_i=1$. The parameter $v_i$ is the rest frame velocity of the $i$-th hyperfine line with respect to that of the main hyperfine component. The values for $\tau_i$ and $v_i$ are taken from \citealt{1964ThaddeusHyperfine,2009PaganiFrequency} and \citealt{2002AhrensSubdoppler} for \nhtd, \nthp\ and \hcn, respectively.
A fit was performed at every position for which the integrated intensity was $\ge3\sigma$. 
The fits results for $T_{ex}$, $v_c$, $\sigma_v$ and $\tau_{tot}$ were recorded only if their values were at least 3 times their errors. The fits had no problem at constraining central velocities (Figure~\ref{fig:l1451-mmvel}) and linewidths (Figure~\ref{fig:l1451-mmsigma}). However, excitation temperatures and opacities were not always well constrained (i.e. the ratio between the parameter value and its error, obtained from the fit, was less than 3), probably due to the degeneracy between excitation temperature and opacity in the optically thin cases. The fits that were successful at constraining the four free parameters resulted on average excitation temperatures of 5.3, 6.1 and 6.4 K  and average total opacities of 3.8, 9.8 and 5.2 for \nhtd, \nthp\ and \hcn, respectively.  The highest intensity hyperfine component in the spectrum of \nhtd, \nthp\ and \hcn\ has an individual opacity corresponding to 0.5, 0.25 and 0.5 of the total opacity, respectively. Thus, the results of the fits indicate that these hyperfine components have opacities values of $\sim$ 2-3. On the other hand, the weakest hyperfine components have an individual opacity of 0.1 of the total opacity for all the molecules in our sample. Hence, these hyperfine components are on average optically thin. 
In the case of molecular spectra with no hyperfine structure we fit the line with a single Gaussian with only $v_c$, $\sigma_v$ and the peak intensity as free parameters (last two panels in Figures~\ref{fig:l1451-mmvel} and~\ref{fig:l1451-mmsigma}).

\subsection{Envelope Mass} \label{sssec:mass}

We can estimate LTE masses from the observed emission of \nhtd\ and \nthp.  These masses are lower limits since we have only interferometric observations in which extended emission is resolved out. However, these molecules trace high density gas, which is compact and likely does not have a significant extended emission component that is larger than 50" or 12000 AU (which is the largest scales our interferometric observation are sensitive to). Furthermore, we use the \nthp\ data cube of this region, published in the recent work by \cite{2016StormCARMA}, for comparison. The observations in that work fully recover emission at all scales. Comparing both spectra, we estimated an average missing flux of about 30\%, calculated over circular regions with a radius from 6" to 20". Emission from \nhtd\ is typically less extended than the emission from \nthp\ and therefore we expect even less missing flux for this molecule (\citealt{2013DanielNitrogen,2014ChitsazzadehPhysical}). \\

First, we calculate the column density, $N_{tot}$, in each pixel. From \cite{2015MangumHow}, column density can be calculated as:

\begin{equation}
\begin{split}
N_{tot}=\frac{8\pi\nu^3}{c^3 A_{ul}}\frac{Q_{rot}(T_{ex})}{g_u}e^{E_u/T_{e}}
\cdot[e^{h\nu/kT_{e}}-1]^{-1}\int \tau_v dv
\label{eq:columndensity} 
\end{split}
\end{equation}
where $\nu$ is the rest frequency of the emission line, $A_{ul}$ is the Einstein A coefficient for the observed transition, $g_u$ and $E_u$ are respectively the statistical weight and energy of the upper level for the rotational transition, and $Q(T_{ex})$ is the rotational partition function of the molecule. Assuming that the emission line has a gaussian profile, we can replace the integral term by $\int \tau_vdv=\tau_{tot} \sqrt{2\pi}\sigma$. The values for $\sigma$, $T_{ex}$ and $\tau_{tot}$ are obtained from the gaussian fits to the spectra. The rest frequency of the main hyperfine line is used as the $\nu$ value. The values for $A_{ul}$ and $g_u$ were taken directly from the CDMS catalog (\citealt{2001MullerCologne, 2005MullerCologne}). The functional form for $Q(T_{ex})$ for linear molecules such as N$_2$H$^+$ depends linearly on $T_{ex}$, on a first order approximation. In the case of NH$_2$D, a slightly asymmetric top molecule, the first order approximation for $Q(T_{ex})$ is proportional to $T_{ex}^{1.5}$ \citep{2015MangumHow}. Thus, in order to obtain $Q(T_{ex})$ we fit the CDMS catalog values for $Q(T_{ex})$ as $a+bT_{ex}$ and $a+bT^{1.5}_{ex}$ for N$_2$H$^+$ and NH$_2$D, respectively \citep{2014ChitsazzadehPhysical}. From these fits we obtain $Q(T_{ex})=4.41+0.74T^{1.5}_{ex}$ for NH$_2$D and $Q(T_{ex})=3.41+4.01 T_{ex}$ for N$_2$H$^+$. \\

As we noted earlier, fits to the molecular line spectra were not always successful at obtaining the total opacity and excitation temperature. Yet, we need estimates of these two parameters to obtain an estimate of the column density (see Eq.~\ref{eq:columndensity}). Thus, for the purpose of obtaining an estimate of the mass from the \nthp\ and \nhtd\ data we fitted the spectra fixing the excitation temperature to a constant value throughout the core. We did two fits for each of these two molecular lines, one with $T_{ex}=10$ K and one with $T_{ex}=6$  K, which result in a lower and upper limit value for the column densities (and therefore masses), respectively. The higher value of $T_{ex}$ was chosen based on the estimate of the kinetic temperature obtained by \cite{2011PinedaEnigmatic}, while the lower value was chosen based on the fits in which the excitation temperature was successfully obtained. We estimate the mass of the envelope within 4200 AU of the central source in order to compare it to the mass estimate from \cite{2011PinedaEnigmatic} of 0.3 M$_{\odot}$, obtained from single-dish millimeter dust continuum emission observations with a resolution of 11". The ranges for the column density peak and mass within 4200 AU for \nthp\ and \nhtd\ are summarized in Table~\ref{tb:l1451-mmmasses}. 

In order to estimate the total mass, that is mainly in the form of molecular hydrogen, we need to assume a value for the fractional abundance of NH$_2$D and N$_2$H$^+$ with respect to H$_2$. We use the values inferred in \citealt{2013DanielNitrogen,2016DanielCollisional} for the B1b core. The B1b core is also in the Perseus molecular cloud and contains the FHSC candidates B1b-N and B1b-S (\citealt{2012PezzutoHerschel,2013HuangProbing,2014HiranoTwo}). Thus, the assumed fractional abundances are $\sim6\cdot10^{-9}$ and  $\sim1.5\cdot10^{-9}$ for NH$_2$D and N$_2$H$^+$, respectively. These values are also within the range of values found in the literature for prestellar and protostellar sources (\citealt{2014ChitsazzadehPhysical, 2010FriesenInitial, 2007ChenOvro, 2004JorgensenMolecular}). The resultant masses are listed in Table~\ref{tb:l1451-mmmasses}.\\

If we assume the core is in virial equilibrium, we can obtain an alternate mass estimate using:
\begin{equation}
M_{vir}=5\sigma_{H_2}^2R/G \label{eq:mvir}
\end{equation}
where a uniform density profile is assumed \citep{1992BertoldiPressure}. Here, $\sigma_{H_2}$ corresponds to the total linewidth (thermal plus non-thermal) of an average molecular gas particle with a mean molecular weight $\mu=2.33$ (assuming gas with 90\% H$_2$ and 10\% He). Following \cite{2007ChenOvro}, we estimate $\sigma_{H_2}$ using the following equation:
\begin{eqnarray}
\sigma_{H_2}^2&=&kT/\mu+\sigma_{NT}^2\\
&=&kT/\mu+(\sigma_{obs}^2-kT/m_{mol}) \nonumber
\end{eqnarray}
where $\sigma_{obs}$ is the total linewidth for the observed molecular line and $m_{mol}$ is the mass of the molecule. 
For $\sigma_{obs}$ we use the average value within the radius used for calculating the virial mass. The virial masses are listed in Table~\ref{tb:l1451-mmmasses_virial}. Uncertainties in the virial masses estimate are due to different choices for the core's density profile. A density profile of $r^{-2}$ (instead of the  constant density assumed in~\ref{eq:mvir}) would reduce the mass estimate by 40\% compared to the estimate using Eq.~\ref{eq:mvir} \citep{2005SchneeDensity}. Table~\ref{tb:l1451-mmmasses_virial} also lists the values for the fractional abundances derived here as the ratio between the masses shown in the third column of Table~\ref{tb:l1451-mmmasses} and the virial masses. The derived ranges for the fractional abundance are consistent with the values taken from the literature, used to obtain the estimate of the total LTE masses shown in the fourth column of Table~\ref{tb:l1451-mmmasses_virial}. 

The envelope mass estimates in Tables~\ref{tb:l1451-mmmasses} and~\ref{tb:l1451-mmmasses_virial} are in good agreement and give an approximate envelope mass (within 4200 AU) of about 1 M$_{\odot}$. This value is higher than, yet consistent with, the estimate given by \cite{2011PinedaEnigmatic} using 1.2 mm
dust continuum emission (0.3 M$_{\odot}$), given the uncertainties 
in both our and their estimate of the mass.

\begin{deluxetable}{cccc}
\tabletypesize{\scriptsize}
\tablewidth{250pt}
\tablecaption{L1451-\textrm{\normalfont mm} LTE Mass estimates \label{tb:l1451-mmmasses}
}
\tablehead{\colhead{Molecule}                       &
           \colhead{N$_{peak}$\tablenotemark{a}}    & 
           \colhead{Mass\tablenotemark{b}}          & 
           \colhead{Total Mass\tablenotemark{c}}  \\ 
           \colhead{}                               &
           \colhead{[cm$^{-2}$]}                   & 
           \colhead{M$_{\odot}$}                    & 
           \colhead{M$_{\odot}$}                    }

\startdata
NH$_2$D &   $1-3.3(\times10^{14})$  &$0.4-1.4(\times10^{-8}$) &$0.7-2.3$\\
N$_2$H$^+$&  $1-1.5(\times10^{13})$ & $1.2-1.6(\times10^{-9}$)& $0.8-1.0$\\
\enddata

\tablecomments{ The lower and upper limits in each case corresponds to the calculation made with the fits that assumed a fixed excitation temperature of 10 K and 6 K, respectively. We assumed a distance of 230 pc.}
\tablenotetext{a}{Column density peak of gas for the particular molecule.} 
\tablenotetext{b}{Total mass of NH$_2$D or N$_2$H$^+$ in envelope within 4200 AU of central source.} 
\tablenotetext{c}{Total gas mass of envelope within 4200 AU of central source. We assumed a NH$_2$D fractional abundance of $6\times10^{-9}$ \citep{2016DanielCollisional} and a N$_2$H$^+$ fractional abundance of $1.5\times10^{-9}$ \citep{2013DanielNitrogen}. }
\end{deluxetable}

\begin{deluxetable}{ccc}
\tabletypesize{\scriptsize}
\tablecaption{L1451-\textrm{\normalfont mm} Virial Mass estimates \label{tb:l1451-mmmasses_virial} }
\tablewidth{250pt}
\tablehead{
\colhead{Molecule} & 
\colhead{Mass$^{Virial}$(H$_2$)\tablenotemark{a}} & 
\colhead{Abundance\tablenotemark{b}}  \\
\colhead{}  &
\colhead{M$_{\odot}$} &
\colhead{}
}
\startdata

NH$_2$D&  $1-1.6$ &$0.4-2.5(\times10^{-8}$)   \\

N$_2$H$^+$ &   $0.7-1.1$  &$1.1-2.3(\times10^{-9}$) \\

\enddata

\tablecomments{These masses are calculated within a radius of $4200$ AU. We assumed a distance of 230 pc.}
\tablenotetext{a}{We assume a kinetic temperature $T=10$ K based on NH$_3$ observations presented by \cite{2011PinedaEnigmatic}}

\tablenotetext{b}{Ratio between the envelope mass for each molecule listed in third column of Table~\ref{tb:l1451-mmmasses} and the virial mass.}

\end{deluxetable}

\subsection{Kinematics}

\subsubsection{Rotation} \label{sssec:kin}

\begin{figure}
\includegraphics[width=0.48\textwidth]{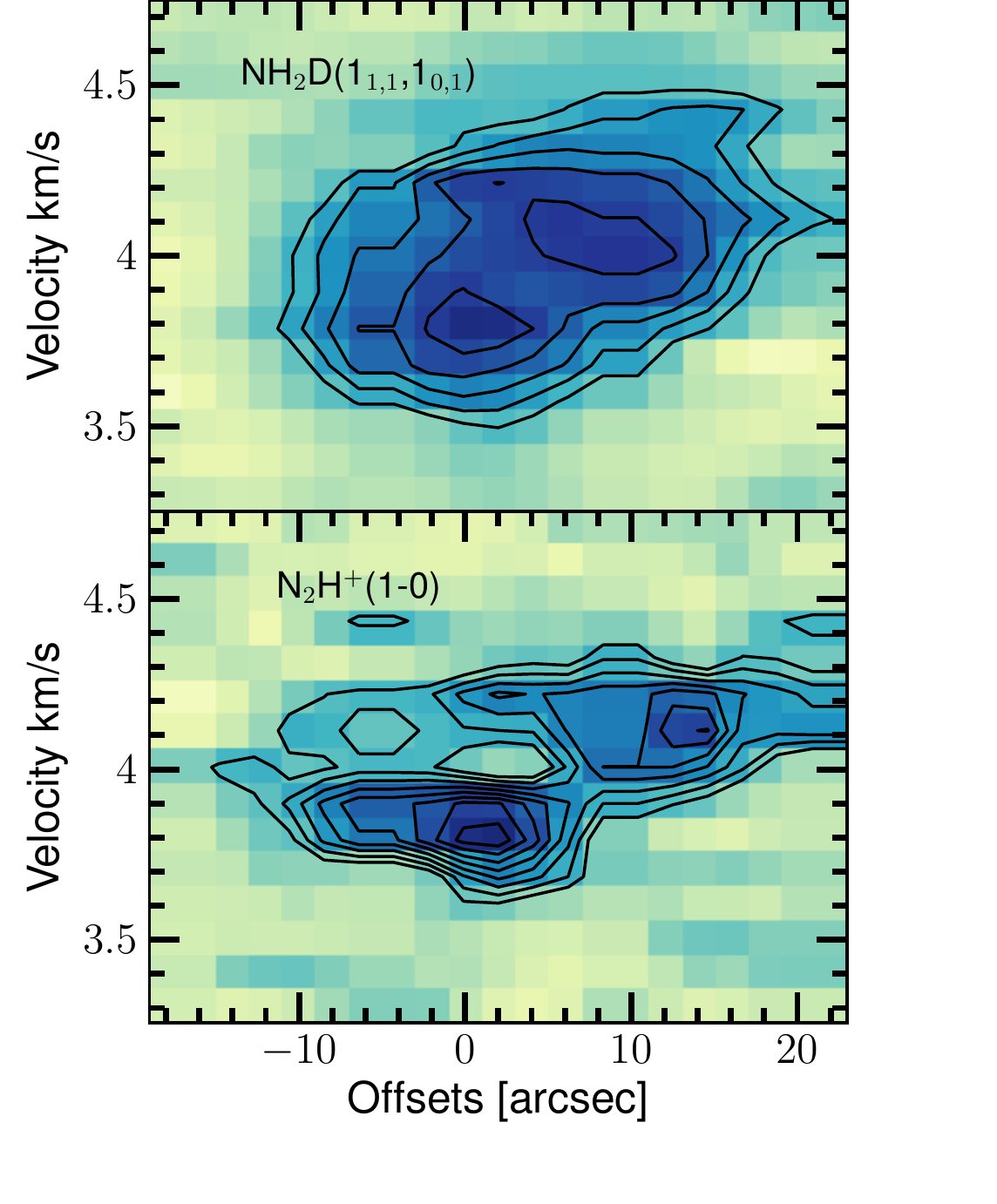}
\caption{L1451-mm position velocity maps of the \nhtd\ (top panel) and \nthp\ (bottom panel) lines along a cut perpendicular to the outflow lobes direction. The cut is shown as a yellow line of 40" in length in Figure~\ref{fig:L1451-mmmom0}. The width of the cut is one pixel corresponding to 2". For \nhtd\ only the strongest hyperfine line is shown, while for \nthp\ the emission corresponds to the isolated hyperfine line.  
Black contours enclose emission starting from 3.8$\sigma$ increasing in steps of 1.2$\sigma$ for both molecules.\label{fig:L1451-mmposvel}}
\end{figure}

A velocity gradient is observed in the velocity maps of \nhtd\ and \nthp\ in Figure~\ref{fig:l1451-mmvel}.  This velocity gradient can also be seen in the position-velocity map of these molecules in Figure~\ref{fig:L1451-mmposvel}, where the cut is made at a direction perpendicular to the outflow, at PA$ = 104^{\circ}$ (indicated by the yellow line in Figure~\ref{fig:L1451-mmmom0}). In order to measure the magnitude and direction of the gradient, we fit the velocity maps using the method described in \cite{1993GoodmanDense}. This method assumes that the velocities change linearly with position. By performing the fit in regions of different sizes, we can study variations on the magnitude and position angle of the gradient, as a function of distance from the continuum peak. For each molecule, we used elliptical regions with an aspect ratio and direction consistent with the morphology of the integrated emission. The values for the aspect ratio and direction of the elliptical regions were taken from the fits to the integrated emission summarized in Table~\ref{tb:l1451-mmproperties}. We set up ellipses with semi-major axis increasing in steps of 7" ($\sim$1610 AU) up to 21" ($\sim$6440 AU) as well as 7"-width annuli,  as the regions for performing the fit. The results of these fits are summarized in Table~\ref{tb:l1451gradients}.

\begin{deluxetable*}{lccc}
\tabletypesize{\scriptsize}
\tablecaption{Velocity Gradient fit results \label{tb:l1451gradients} }
\tablewidth{450pt}
\tablehead{
\colhead{Molecule}&
\colhead{Gradient} & 
\colhead{(PA$_{grad}$-PA$_{out}$)\tablenotemark{a}}&
\colhead{Distance\tablenotemark{b}}  \\
\colhead{}&
\colhead{[\kms\ pc$^{-1}$]}&
\colhead{[$^{\circ}$]}&
\colhead{["]}
}

\startdata

NH$_2$D & $10.1\pm1.3$& $80\pm9$&$\leqslant7$ \\
& $8.7\pm0.5/8.4\pm0.4$&$79\pm4/83\pm4$&$\leqslant14/(7,14]$ \\
& $7.6\pm0.4/7.1\pm0.3$&$76\pm4/82\pm4$&$\leqslant21/(14,21]$ \\
N$_2$H$^+$&  $9.9\pm1.7$& $38\pm9$& $\leqslant7$  \\
& $5.8\pm0.5/5.7\pm0.3$ &$87\pm6/93\pm4$&$\leqslant14/(7,14]$\\
& $5.1\pm0.2/5.0\pm0.2$&$89\pm4/93\pm3$ &$\leqslant21/(14,21]$ 

\enddata
\tablecomments{left/right values correspond to fits in the respective regions indicated in column 4.}
\tablenotetext{a}{Gradient position angle minus outflow position angle where PA$_{out}=14^{\circ}$.}
\tablenotetext{b}{Length of the semi-major axis of the elliptical region where the fit is performed (left). Lower and upper limit to the semi-major axis of the annulus where the fit is performed (right). The elliptical regions are centered at the continuum peak.\\} 
\end{deluxetable*}

For the \nhtd, the magnitude of the gradient increases slightly from 7.1 $\pm$ 0.3 \kms\ pc$^{-1}$ near the edge of the envelope, to 10.1 $\pm$ 1.3 \kms\ pc$^{-1}$ at the center. The direction of this gradient is fairly constant as a function of distance from the center. The average difference between the outflow and velocity gradient direction for the annuli regions is $82\degr\pm1\degr$. The average systemic velocity obtained in the fits is 3.9 \kms. 

The magnitude of the velocity gradient for the \nthp\ also shows a slight increase towards the center. However, the inner 7" region (i.e., the region that mostly shows blue velocities) is likely to be tracing infall motions (see Infall section bellow). Thus, the direction of the gradient in this region is not consistent with the values obtained using larger scales. Excluding the inner region, the direction of the gradient remains close to perpendicular to the outflow with a difference between the outflow and gradient direction for the annuli regions of $93\degr$. The average systemic velocity obtained for \nthp\ is 3.9 \kms, the same as that obtained using the \nhtd\ velocity map. 
The gradients of both \nhtd\ and \nthp\ and their direction using the larger region ($\leqslant$21") are consistent with the values reported in \cite{2011PinedaEnigmatic}. 
The trend of these two tracers, of increasing gradient magnitudes with decreasing distance from the center, is consistent with rotation that increases towards the center. We explore this further in Section 4.3.3.

\hcn, \hcop\ and \cs\ do not show clear rotation signatures in the velocity maps shown in Figure~\ref{fig:l1451-mmvel}. Fits to these velocity maps result in highly uncertain gradients with magnitudes less than 1 to 2 km s$^{-1}$ pc$^{-1}$.

\subsubsection{Infall} 

The \nthp\ velocity map in Figure~\ref{fig:l1451-mmvel} shows a blue region in the center, close to the continuum peak position. A concentration of blue velocities at the center of a velocity map is known as a ``blue bulge''  and it is a signature of infall that is produced when the molecular line emission is optically thick (\citealt{1994WalkerSpectroscopic,2006WilliamsHigh}). When the transition is optically thick most of the redshifted emission comes only from the near-side outer layers of the envelope which has a lower excitation temperature compared to the inner layers (behind the central source) where most of the blueshifted emission comes from. This results in an asymmetric profile with higher intensity in the blueshifted side of the line (see \citealt{1999EvansPhysical}). The asymmetric profile can contain one or two peaks depending on how the infall velocity and excitation temperature vary with radius. In a velocity map (or first-moment map) this asymmetry results in a concentration of blueshifted velocities at the center of the core (a.k.a., a blue bulge).

\begin{figure}
\includegraphics[width=0.5\textwidth]{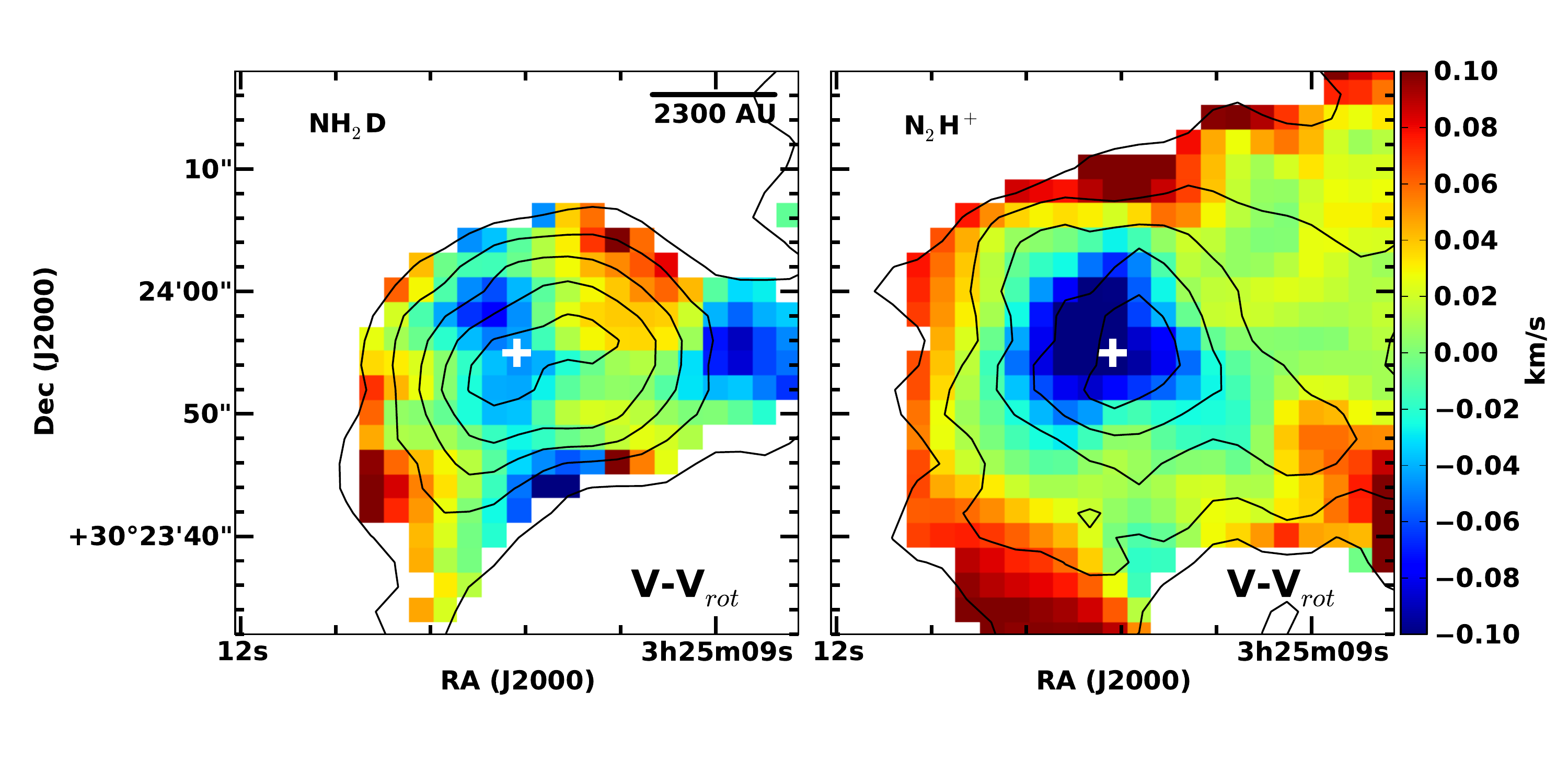}
\caption{L1451-mm gradient-subtracted velocity maps for \nhtd\ and \nthp. The magnitude of the subtracted gradient in each case is shown in Table~\ref{tb:l1451gradients}, and corresponds to the elliptical region with semi-major axis $\leqslant21"$. The black contours are the same as in Figure ~\ref{fig:L1451-mmmom0}. The cross marks the position of the continuum. \label{fig:L1451-mmsub}}
\end{figure}

In order to search for evidence of infall using the ``blue bulge'' signature in the \nthp, we first subtracted the rotation gradient from the velocity map of this molecular line, following the procedure in \cite{2006WilliamsHigh}. The magnitude and direction of the subtracted gradient was calculated as explained in Section 4.2.1, using an elliptical region with semi-major axis of 21" (i.e excluding the extended emission tails). We applied the same procedure to \nhtd, to test the presence of a blue bulge in the map of this molecule as well. Figure~\ref{fig:L1451-mmsub} shows the central velocity maps for \nhtd\ and \nthp\ after subtracting the rotation gradient. The \nthp\ shows a clear blue bulge at the center of the core. The \nhtd\ also shows blue velocities at the center, but the blue bulge signature is not as pronounced as that of the \nthp\ map. This is probably because the opacity of the \nhtd\ line is less than that of \nthp.

\begin{figure}
\includegraphics[width=0.5\textwidth]{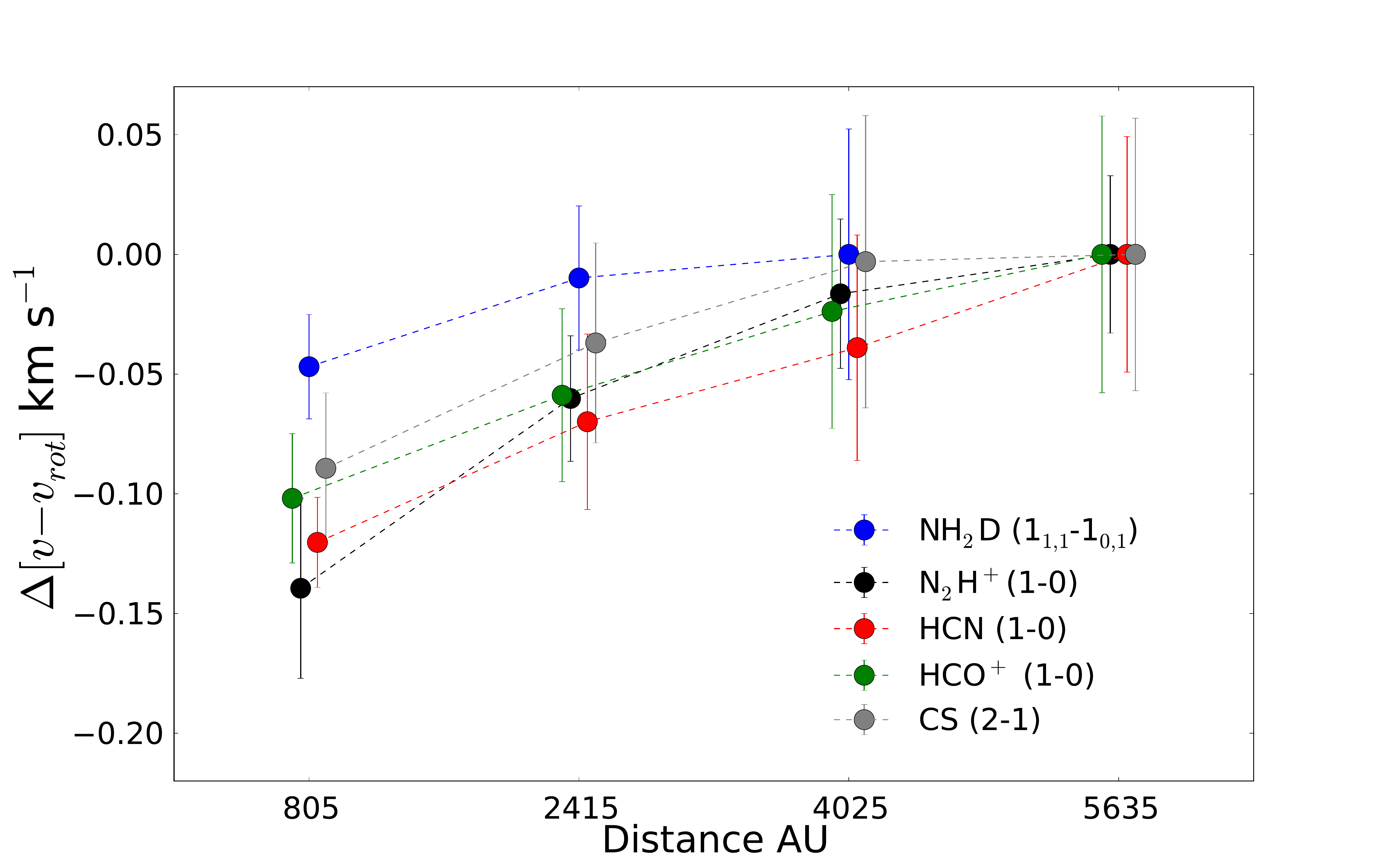}
\caption{Velocity shift toward the center of L1451-mm as traced by each molecular line. For \nhtd\ and \nthp\ the shift in velocity was calculated from the central velocity map {\it after} subtracting the velocity gradient (Figure~\ref{fig:L1451-mmsub}), while for \hcn, \hcop\ and \cs\ the shift was calculated directly from the central velocity map in Figure~\ref{fig:l1451-mmvel} (i.e., no gradient was subtracted). The velocity shift at each distance corresponds to the average velocity over an elliptical ring that follow the shape of the integrated intensity map, with a width of 7". The values are scaled such that the average velocity at the outermost radii is zero. The errors are the standard deviation of the average.\\ 
\label{fig:L1451-mminfallprofile}}
\end{figure}

Following \cite{2006WilliamsHigh}, we can examine the blueward velocity shift as a function of the distance from the continuum peak, using the velocity maps for which the velocity gradient has been subtracted (Figure~\ref{fig:L1451-mmsub}). For this, we calculated the average velocity in the gradient-subtracted maps in annuli of 7" width, up to 28" away from the continuum peak. The orientation and aspect ratio of the elliptical rings were taken from the fits to the integrated emission summarized in Table~\ref{tb:l1451-mmproperties}. The velocity shift was then obtained by subtracting the average velocity in each $7"$-wide ring to the average velocity of the ring furthest from the continuum source (see Figure~\ref{fig:L1451-mminfallprofile}). The size of the envelope traced by these two species is different and therefore the radius at which the velocity shift is set to zero is smaller for \nhtd\ compared to that of \nthp. Similarly, we also searched for infall evidence in the other three molecules on our sample. In this case, we did not use a gradient-subtracted map (as the velocity maps of these molecular lines do not show a rotation signature) and the velocity shifts were calculated using the central velocity maps in Figure~\ref{fig:l1451-mmvel}. Figure ~\ref{fig:L1451-mminfallprofile} shows the velocity shift as a function of the distance from the continuum peak for all the molecules in our sample. All the species show a blueward velocity shift that increases with decreasing radius, with \nthp\ showing the larger difference between the edge and the center of the envelope.\\

\begin{figure}
\includegraphics[width=0.5\textwidth]{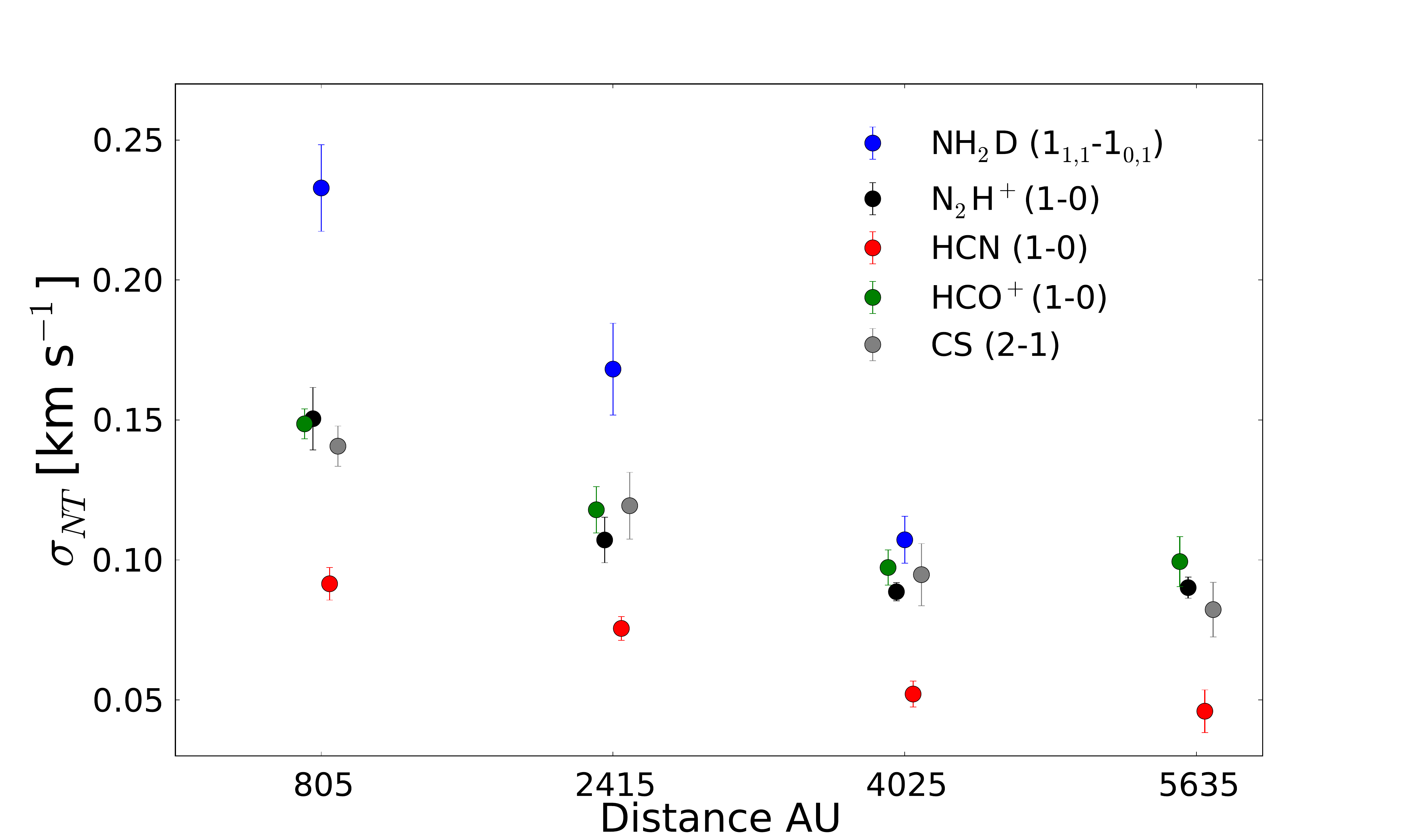}
\caption{L1451-mm non-thermal linewidths obtained at different distances from the continuum. These values were obtained over the same regions as the velocity shifts shown in Figure~\ref{fig:L1451-mminfallprofile}, using the total (observed) linewidth maps shown in Figure~\ref{fig:l1451-mmsigma}, and assuming a constant temperature of 10 K for estimating the thermal component of the linewidth. \label{fig:L1451-mmwitdthprofile}}
\end{figure}

Figure~\ref{fig:l1451-mmsigma} shows linewidths maps for all of the observed molecular lines. The broad linewidth region close to the center show widths that would imply temperatures $\gtrsim40-70$ K if they were only caused by thermal motions. Such high gas temperatures are highly unlikely at the distances measured here ($\gtrsim$ 1000 AU) for Class 0 sources (\citealt{1993TerebeyContribution,2014MauryFirst}). Thus, these broad linewidths regions are most likely caused by non-thermal motions such as rotation/infall and outflow motions. Figure~\ref{fig:L1451-mmwitdthprofile} shows average non-thermal linewidths as a function of the distance from the continuum source, the average is calculated over the same regions as the infall velocities shown in Figure~\ref{fig:L1451-mminfallprofile}, using the values from Figure~\ref{fig:l1451-mmsigma} and assuming a thermal width calculated at a temperature of 10 K.
The non-thermal widths increase toward the center for all molecular line maps, particularly for \nhtd\ which is the only molecule showing non-thermal widths larger than the sound speed in the inner $\sim$1610 AU region ($c_s=0.19$ \kms\ at $T=10$ K). In the case of the \nthp\ these larger linewidths are mainly caused by infall motions which we trace using the double peak profile in the spectra and the blue bulge in the central velocity map. There could also be contribution from rotational motions. The peak of the linewidth map in the \nhtd\ is spatially coincident with the direction of the blue outflow lobe. However since the large scale increased linewidth region seems to be aligned with the semi-minor axis of the envelope, these large linewidths could also be produced by infall/rotational motions. \\

\begin{figure*} 
\centerline{\includegraphics[height=2in]{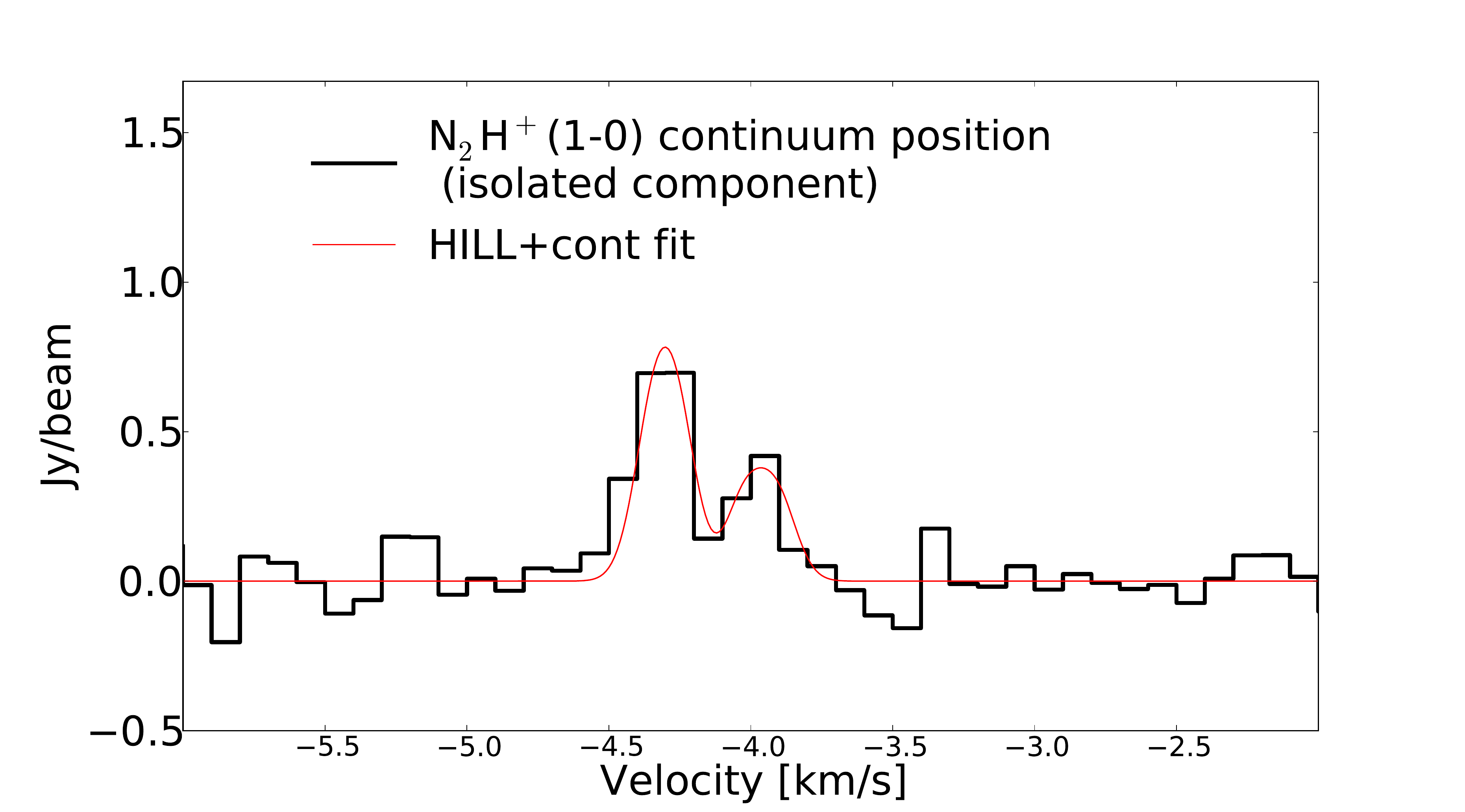}\includegraphics[height=2in]{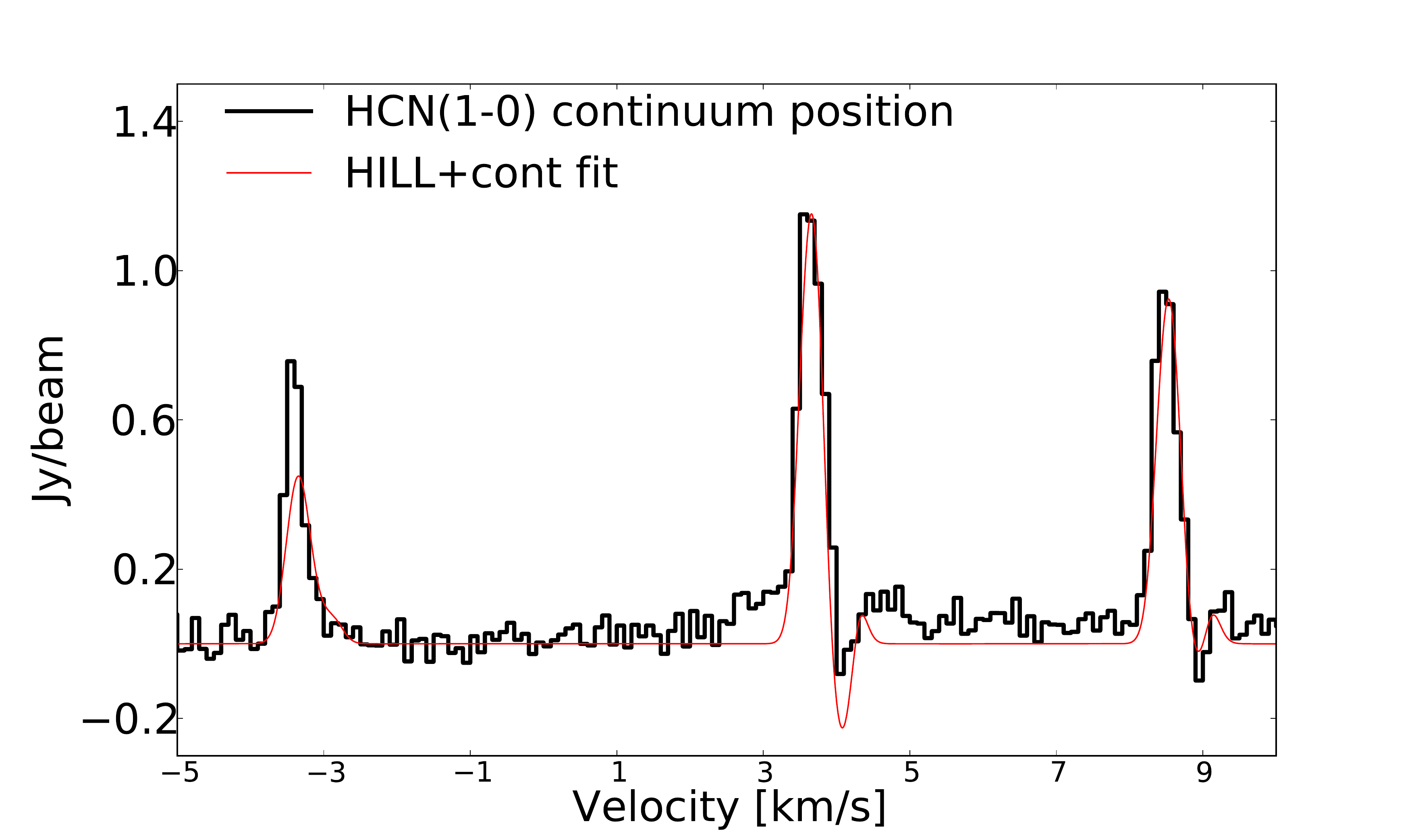}}
\caption{\label{fig:infall_profile_l1451-mm} 
{{\it Left:} L1451-mm \nthp\ spectrum at the position of the continuum source. Only the isolated component is shown. The spectrum was fitted using the HILL model with additional absorption from a continuum source. The infall velocity from the fit (in red) is  0.17 $\pm\ $ 0.05 \kms. }  
{{\it Right:} L1451-mm \hcn\ spectrum at the position of the continuum source. The three hyperfine components are shown. The two most intense hyperfine components show an inverse p-cygni profile. The spectrum was fitted using the HILL model with additional absorption from a continuum source. The infall velocity from the fit (in red) is 0.17 $\pm\ $ 0.04 \kms. 
}}
\end{figure*}

As discussed earlier, infall motions in L1451-mm are also revealed through the detection of an inverse p-cygni profile in \hcn\ and a double peaked, blue asymmetric profile of the isolated component of \nthp, both at the position of the continuum source. Figure \ref{fig:infall_profile_l1451-mm} shows these two spectra. The absorption of the continuum is clearly seen in the two more intense \hcn\ hyperfine components. We fit both spectra separately, using the HILL model \citep{2005deVriesMolecular} with the addition of a continuum source following the procedure in \cite{2001diFrancescoInfall}. In the HILL model the excitation temperatures varies linearly with the optical depth, between the outer layer and the center of the core, while for the two-layers model \citep{1996MyersSimple} the excitation temperature is a step function. The parameters of the fitting function are the systemic velocity $v_{sys}$, infall velocity $v_{inf}$, excitation temperature at the center $T_{center}$, excitation temperature of the outer layer $T_{outer layer}$, linewidth of the line $\sigma_v$, the line opacity $\tau$ (total opacity in the case of \hcn\ and the opacity of the isolated hyperfine line in the case of \nthp), the temperature of the continuum $T_c$ and the fraction of the beam that is occupied by the continuum $\phi$. To reduce the amount of free parameters we fixed the values of $v_{sys}$, $T_c$, $\phi$ and $T_{outer layer}$. For the systemic velocity, we used 3.9\kms\ which is the value obtained after fitting the velocity gradient of both, the \nhtd\ and \nthp\ velocity maps (see Section 4.3.1). We followed \cite{2012PinedaFirst} and fixed $T_c$ to the radiation temperature that matched the peak of the 3mm continuum flux and $T_{outer layer}$ to a conservative lower limit of 3 K. The value for $\phi$ was fixed to 0.3, which corresponds to a 3mm source of $0.3\times6"\times230=414$ AU in size. As discussed in \cite{2012PinedaFirst}, this value for $\phi$ is arbitrary, but the chosen value does not affect the results for the infall velocity and linewidth. If a higher value were to be used for $\phi$, this would result in a higher value for $T_{center}$ and a lower value for $\tau$. 
For the \nthp\ profile, the fit was performed for the isolated hyperfine only (Figure~\ref{fig:infall_profile_l1451-mm}). This is because the isolated hyperfine component in our observations does not follow exactly the intensity ratio (with respect to the other hyperfines components) expected from LTE conditions. In this case, the isolated hyperfine is brighter. Therefore, the best fit to the entire spectrum is not the best fit for the isolated hyperfine which is the one showing the infall signature more clearly. For \hcn\ the fit was made simultaneously to the three hyperfine lines (Figure~\ref{fig:infall_profile_l1451-mm}). In this case, the fit underestimated the intensity for the weakest hyperfine. Similar to the \nthp\ spectrum, the weakest hyperfine in the \hcn\ is also brighter than expected from LTE conditions. This anomaly has been seen before towards low mass cores (\citealt{2007DanielExcitation,2014ChitsazzadehPhysical,2012LoughnaneObservations}). The fixed and free parameters as well as the results from the fits are summarized in Table~\ref{tb:l1451-mmm_hillfit}. The derived infall velocity in the central part of the envelope ($\sim$0.17\kms) is consistent with observations at 1000 AU scales of the FHSC candidate Cha-MMS1 \citep{2013TsitaliDynamical}, the young Class 0 source IRAM 04191 \citep{2002BellochMolecular}, surveys of Class 0 sources \citep{2013SchneeCorrelating} as well as simulations of the first core stage \citep{2015TomidaRadiation}.

The velocity map of \hcn\ and \hcop\ show a concentration of blue velocities towards the center, albeit not as pronounced as that seen in the \nthp\ velocity map. Since the \hcn\ spectrum shows evidence of infall, and the region with large linewidth does not necessarily align with the outflow, it is likely that the blue velocities near the center are related to infall motions. It is unclear if the \hcop\ and \cs\ emission are tracing some type of ordered motion or outflow/envelope interactions (see Section 5). 

\begin{deluxetable*}{lccccccccccrl}
\tabletypesize{\scriptsize}
\tablecaption{L1451-\textrm{\normalfont mm} HILL+Continuum fit to \nthp\ and \hcn\ central spectra  \label{tb:l1451-mmm_hillfit} }
\tablewidth{450pt}
\tablehead{
\colhead{Mol}&
\colhead{$v_{sys}$\tablenotemark{a}} & 
\colhead{$T_{continuum}$\tablenotemark{a}}  & 
\colhead{$T_{outer layer}$\tablenotemark{a}}& 
\colhead{$\phi$\tablenotemark{a}}&
\colhead{ $T_{center}$\tablenotemark{b}}&
\colhead{$v_{inf}$\tablenotemark{b}}&
\colhead{$\sigma_v$\tablenotemark{b}}&
\colhead{$\tau$\tablenotemark{b,c}}\\
\colhead{}&
\colhead{\kms}&
\colhead{K}&
\colhead{K}&
\colhead{}&
\colhead{K}&
\colhead{\kms}&
\colhead{\kms}
}
\startdata

N$_2$H$^+$(1-0)&   3.9  &19&3&0.3&21 $\pm$ 1 &0.17 $\pm$ 0.05 &0.08 $\pm$ 0.01&0.78 $\pm$ 0.07\\
HCN(1-0)&   3.9  &19&3&0.3&15 $\pm$ 2 &0.17 $\pm$ 0.04 &0.16 $\pm$ 0.01 &5.0 $\pm$ 0.5\\

\enddata
\tablenotetext{a}{Fixed parameters.}
\tablenotetext{b}{Free parameters.}
\tablenotetext{c}{Line opacity $\tau$ corresponds to the total opacity in the case of \hcn\ and to the opacity of the isolated hyperfine line in the case of \nthp.}

\end{deluxetable*}

\subsubsection{Infalling and rotating envelope model} \label{sssec:model}

Recent simulations that include the effects of magnetic fields, rotation \citep{2014BateCollapse,2015TsukamotoEffects,2015TomidaRadiation} and turbulence \citep{2011MatsumotoProtostellar,2013JoosInfluence} have consistently predicted the formation of the FHSC. The envelope flattens during the collapse and it can develop an oblate or in some cases filamentary shape, with the minor axis aligned with the magnetic field direction \citep{2011MatsumotoProtostellar}.  Infall motions dominate over rotation in the envelope. At scales larger than few hundreds AU the infall velocity ($v_{\inf}$) can be approximated to have a $r^{-0.5}$ dependence, and the rotational velocity ($v_{rot}$) to have a $r^{-1}$ dependence, corresponding to conservation of angular momentum (\citealt{2015TomidaRadiation}).  At later times, Keplerian rotation ($v_{rot}\propto r^{-0.5}$) develops at scales of a few AU, forming a circumstellar disk that is expected to grow during the Class 0 phase (\citealt{2014MachidaConditions, 2013TomidaRadiation, 2014BateCollapse}). A circumstellar disk, corresponding to a rotationally supported FHSC, can also exist before protostar formation  \citep{2015TomidaRadiation}. In this later case the rotational profile is not necessarily Keplerian because the gas pressure remains significant.

High resolution studies, using high density tracers, have found Keplerian disks in the 50-150 AU range for a few Class 0 sources (\citealt{2012TobinNature,2013MurilloKeplerian,2014LeeAlma,2014LindbergAlma}). The gas infall and rotation velocity profiles in the envelope surrounding such disks are observed to be consistent with $v_{inf}\propto r^{-0.5}$ and $v_{rot}\propto r^{-1}$, respectively (\citealt{2013MurilloKeplerian,2014LeeAlma,2014OhashiFormation}). Furthermore, several Class 0/I protostar have been inferred to have disk sizes smaller than 1000 AU \citep{2015YenObservations} or even 10-45 AU (\citealt{2014MaretFirst,2014OyaSubstellar,2015YenNokeplerian}). Thus, we expect the envelope kinematics of a FHSC or a young Class 0 protostar, at scales larger than few 100 AU up to at least 5000 AU \citep{2013BellocheObservation}, to be consistent with infall in which the angular momentum is conserved. \\

We use a toy model to probe the envelope kinematics by comparing the model with our observations. In particular, we contrast the model with the position velocity map of the \nhtd\ line. We use only the \nhtd\ emission since it seems to be
less shaped by optical depth effects, and therefore is a better comparison with our toy model that uses an optically thin approximation. As discussed earlier, the emission from \nthp\ shows a blue asymmetric double peak profile at the continuum position, which can be reproduced by a function that considers continuum absorption and self-absorbed redshifted emission (Figure~\ref{fig:infall_profile_l1451-mm}).  
 
\paragraph{Kinematics of the model:} The model considers that the envelope is composed by particles that follow ballistic orbits due to the gravitational influence of a central mass. The trajectories and velocities of the particles along these orbits were calculated by \cite{1976UlrichInfall} and \cite{2009MendozaAnalytic}. Here, we briefly describe the assumptions and equations that lead to these orbits. We use spherical coordinates.\\

The orbits are defined such that when particles are at a distance $r_0$ from the center, they have the same azimuthal angular velocity $\dot{\phi}_0$,  while $\dot{\theta}_0=0$, i.e. they follow solid body rotation around the z-axis. The parameter $r_0$ corresponds in the model to the radius of the envelope. 

Along each orbit the specific energy E and specific angular momentum  $j$ are conserved. The specific angular momentum $j$ at $r_0$ is given by:
\begin{equation}
j^2(r=r_0)=\dot{\phi}_0^2r_0^4\sin^2{\theta_0}=j_0^2\sin^2{\theta_0}
\label{eq:hatr0}
\end{equation}
and thus, the specific energy for each orbit is given by:
\begin{equation}
E=\frac{v_r^2}{2}+\frac{j_0^2\sin^2{\theta_0}}{2r^2}-\frac{GM}{r}
\end{equation}
where $v_r$ is the radial velocity and $M$ the central mass that dominates the dynamics. Following \cite{1976UlrichInfall}, we assume E=0 which results in parabolic orbits. The radial velocity $v_r$ of a particle along these orbits increases toward the center until a maximum at $r_{v^{max}_r}$, which is given by:
\begin{equation}
r_{v^{max}_r}=\frac{j_0^2\sin^2{\theta_0}}{GM}
\label{eq:r_vinfmax}
\end{equation}

After this radius, $v_r$ rapidly decreases reaching zero at the centrifugal barrier (CB) radius given by $r_{v^{max}_r}/2$. The total non radial velocity in the plane of the orbit increases also toward the center as $\frac{j_0}{r}$. In the midplane, i.e $\theta_0=90^{\circ}$, Eq.~\ref{eq:r_vinfmax} becomes:  
\begin{equation}
r_u=\frac{j_0^2}{GM}
\label{eq:r_u}
\end{equation}
which is the disk formation radius as defined by \cite{1976UlrichInfall}. 

The velocities of the particle along the orbit can be calculated solving the energy equation. The velocities of a particle along its orbit are given by:
\begin{eqnarray}
v_r(r,\theta,\phi)&=&-\Big(\frac{2GM}{r}-\frac{j_0^2\sin^2{\theta_0}}{r^2}\Big)^{\frac{1}{2}} \\ 
v_{\theta}(r,\theta,\phi)&=&\frac{j_0\sin{\theta_0}}{r\sin{\theta}}(\cos^2{\theta_0}-\cos^2{\theta})^{\frac{1}{2}}\\
v_{\phi}(r,\theta,\phi)&=&\frac{j_0}{r}\frac{\sin^2{\theta_0}}{\sin{\theta}}
\label{eq:velocities}
\end{eqnarray}
Thus, we can obtain the velocity along each axis for any position $(r,\theta,\phi)$ within the sphere of radius $r_0$, provided that we have the values for $\theta_0$, $\phi_0$, $M$ and $h_0$. The values for $\theta_0$ and $\phi_0$ (the polar and azimuthal angles when the particle is at a distance $r_0$ from the center) can be obtained for every position $(r, \theta,\phi)$ by using the equation of motion  $r(r,\theta,\phi)$ plus geometrical relations (see more details in \citealt{1976UlrichInfall,2009MendozaAnalytic}). Given a central mass M, the value for $j_0$ can be set by choosing a value for $r_u$ (equation~\ref{eq:r_u}). Thus, the envelope kinematics are defined by two parameters: the central mass $M$ and $r_u$. The latest corresponds to twice the value of the CB at the midplane. We will identify models by their CB value instead of their $r_u$ value. 

\paragraph{Spectra and data cube generation:} Given a coordinate position in the sky $(\alpha,\delta)$ we simulate a line of sight that crosses the spherical envelope at that position. The line of sight can have an inclination $i$ defined such that if $i=90^{\circ}$ the observer is looking in the direction of the x-axis. For each position $(r,\theta,\phi)$ within a given line of sight we calculate the velocity along each axis $(v_r,v_{\theta},v_{\phi})$ using (\ref{eq:velocities}). The projection of $(v_r,v_{\theta},v_{\phi})$ onto the line of sight $v_{los}$ is then given by:

\begin{eqnarray}
\label{eq:vlos}
v_{los}(r,\theta,\phi,i)=v_r(r,\theta,\phi)(\sin{\theta}\cos{\phi}\sin{i}+\cos{\theta}\cos{i}) \nonumber \\
+v_{\theta}(r,\theta,\phi)(\cos{\theta}\cos{\phi}\sin{i}-\sin{\theta}\cos{i}) \nonumber \\   
+v_{\phi}(r,\theta,\phi)(-\sin{\phi}\sin{i}) \nonumber \\
\end{eqnarray}

For each $(r, \theta,\phi)$ along a line of sight we set up a Gaussian centered at a velocity given by $v_{los}$. A Gaussian was calculated every 10 AU along the line of sight. The width of the Gaussian was set to be produced only by thermal motions at a temperature of 10 K. The amplitude of the Gaussian was set to be proportional to the density at that position assuming a density profile given by $r^{-2}$. The final spectrum along a line of sight is the sum of all the Gaussians along the line of sight. 
We set up a grid of lines of sight that cross the spherical envelope with a separation of 55 AU between them. In order to match our observations, we sum all the spectra produced by lines of sight that fall within a single pixel in our observations. Given the pixel size of our observations and the distance to the source, every observed spectrum (corresponding to one pixel in the cut) is modeled with the sum of 81 spectra/lines of sight. 

\begin{figure*}
\includegraphics[width=1\textwidth]{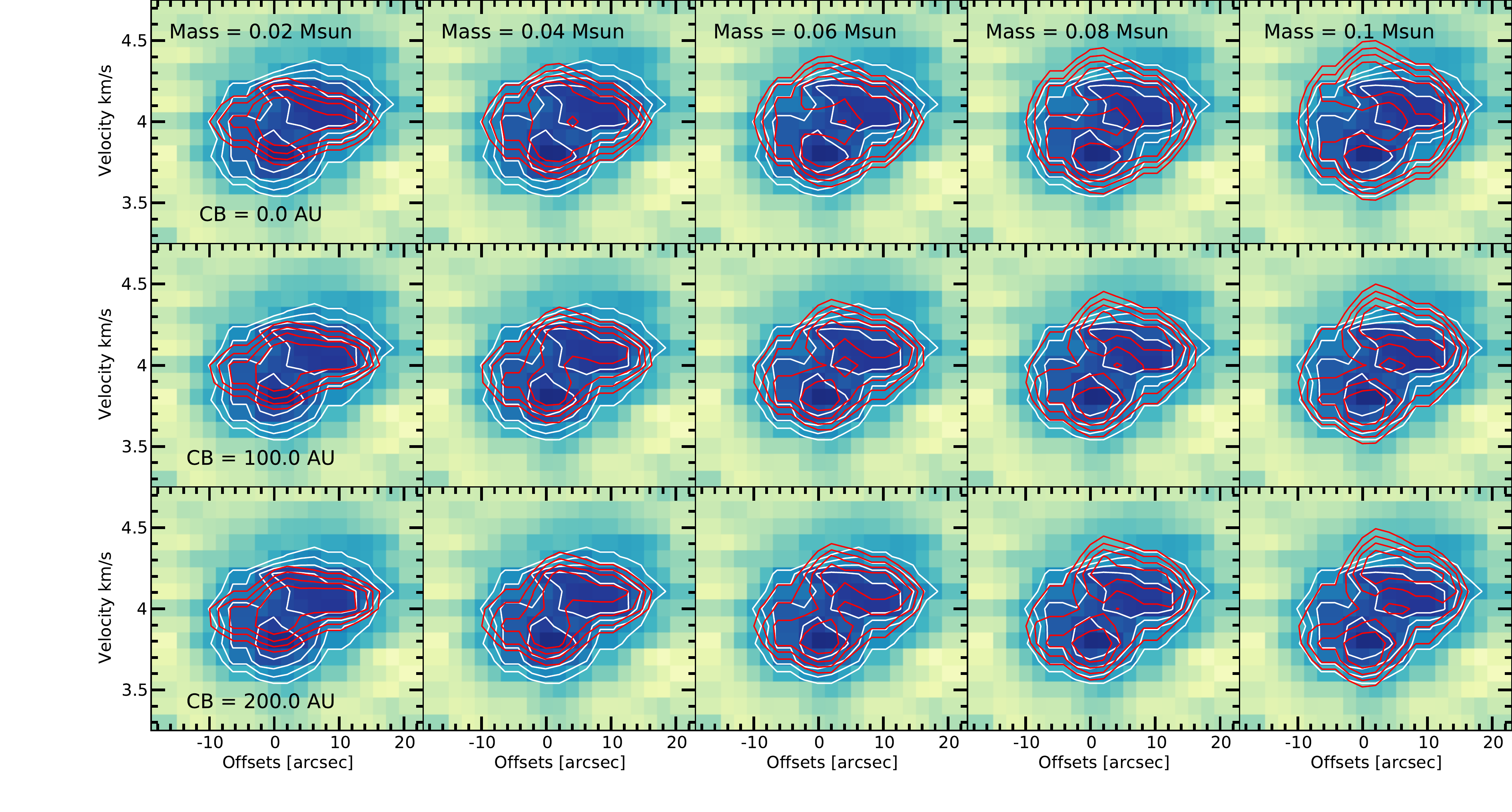}
\caption{L1451-mm position velocity map for \nhtd\ at a direction perpendicular to the outflow (yellow line in Figure~\ref{fig:L1451-mmmom0}). Color map and white contours show the observed data. Red contours show synthetic observations of a model considering particles that, under the influence of a central mass, are infalling and rotating while conserving their angular momentum. Both contours start at 5$\sigma$ and increase in steps of 3.8$\sigma$, where $\sigma$ corresponds to the rms of the observed p-v map. The two parameters of the model are the Centrifugal Barrier (CB) radius at which the infall velocity reaches zero (rows) and the central mass (columns). The best agreement is for a central mass between 0.04 M$_{\odot}$ and 0.08 M$_{\odot}$ with the lowest difference between the model and data for a mass of 0.06 M$_{\odot}$ and a CB $=100$ AU.  \label{fig:l1451vpmodel}}
\end{figure*}

\paragraph{Position velocity map:} To better compare the model to our interferometric observations, once we have a spectrum for each pixel of our observed map (i.e. a data cube for the model) we performed synthetic observations of the model. We use the software MIRIAD to make synthetic visibilities of the model using the uv-sampling of the observed \nhtd\ data with the task uvmodel. Then, we apply the same imaging procedures used for the observed \nhtd\ cube. We construct the final position velocity map with a cut along the equator of the spherical core. To compare the kinematics between the observed and modeled data, for each spatial position along the p-v cut (i.e., for each column in the p-v diagrams shown in Figure~\ref{fig:l1451vpmodel}) we matched the intensity peak value of the observed \nhtd\ spectrum with the modeled one. Since for each position along the p-v cut we are basically comparing the velocities at which the maximum intensity is reached and not the absolute intensities (we did not include radiative transfer), this scaling does not affect the results.\\

Figure \ref{fig:l1451vpmodel} shows simulated position velocity maps for \nhtd\ assuming different values for the central mass and centrifugal barrier radius. Red contours correspond to the model while color maps and white contours correspond to the observed \nhtd\ position velocity map at a cut perpendicular to the direction of the outflow. In these models, we defined the envelope to have a radius of 4000 AU and to be oriented edge-on with respect to the observer. We see from Figure~\ref{fig:l1451vpmodel} that increasing the mass at a constant CB radius makes the maximum blue and red velocities to increase. On the other hand, increasing the CB radius at a constant mass results in a larger offset between the positions where the maximum red and blue velocities are reached, as well as in a slight decrease in the maximum value of the peak red and blue velocities. Comparing the observed p-v diagram with the model using different parameters, we see that the model that produces the least residuals (i.e., our  ``best fit'' model) is the one with a central mass of 0.06 M$_{\odot}$ and CB $=100$ AU. The residuals for models with a central mass that that differ by 0.02 M$_{\odot}$ or less from that of the best fit model  
do not change by more than 50\%. Hence, we argue that based on our model we can constrain the mass of the central object to be 0.4 M$_{\odot}$ to 0.08 M$_{\odot}$ (with a preferred estimate of 0.06 M$_{\odot}$). Similarly, changing the CB value by more or less than 100 AU does not change the residuals (compared to the best fit of CB $=100$ AU) by more than 50\%. However, in the CB case we can constrain the value of the CB to be about 100 AU using other information. We are certain that the CB is less than 200 AU since     
dust continuum observations at higher angular resolution favor a disk with a size
of $\lesssim$ 100 AU (\citealt{2011PinedaEnigmatic,2015TobinSub}). In addition, it is clear from Figure~\ref{fig:l1451vpmodel} that the pure infall case (i.e., CB $=0$) does not reproduce the observations as the data show that the maximum red and blue velocities are not at the same offset position (as it is in the pure infall model). 
Due to our simple assumptions there are additional sources of uncertainty on our central mass estimations. On one hand, we could be overestimating the central mass since our model does not take into consideration the core's self-gravity and we only assume an edge-on configuration. In the later case, it is because in our simple toy model the kinematics are more dominated by infall outside the midplane. In the extreme case of $i=0^{\circ}$ the model would look like the pure infall case. On the other hand, we may be underestimating the central mass due to our choice of the density profile and the slowing down of the infall velocity due to magnetic fields (e.g. \citealt{2011LiNonideal}). Our choice of density profile ($r^{-2}$) at 1000 AU scales agrees with recent simulations of the first core phase (\citealt{2014BateCollapse,2015TomidaRadiation}), and has been successfully used in the literature to model young Class 0 protostars and FHSC candidates (\citealt{2013TsitaliDynamical,2015YenNo}). A more flattened density profile, also observed in some sources (\citealt{2008EnochMass}), reduces the emission from lines of sights near the center, where the highest velocities are found, and thus a higher central mass is needed to reproduce the observed velocities. We compare our observations with a model with an extremely flat density profile of $r^{-1}$ and find that the best fit is found for a central mass of 0.08 M$_{\odot}$. It is also possible to underestimate the central mass by factors of 2 to 4 if one assumes free-fall infall velocities, because magnetic fields cause infall speeds to slowdown, according to studies by \citealt{2015TomidaRadiation} and \citealt{2014OhashiFormation}.

\section{Discussion}

\subsection{Molecular Emission and kinematics}

The integrated intensity of all the molecules in our sample is centrally peaked and this peak also coincides with the 3mm continuum position, within one synthesized beam size (Figure~\ref{fig:L1451-mmmom0}). This morphology, particularly in the \nthp, is not common in the samples of Class 0 protostars in \cite{2007ChenOvro}  and \cite{2011TobinComplex}. There are several cases in which the protostar is seen at the edge of the core traced by the \nthp\ emission, including sources with a close to edge-on configuration. A drop in the abundance of  N$_2$H$^+$ close to the position of the protostar is an expected outcome after protostar formation, since the temperature of the gas in the inner envelope rises over 20 K and the CO is released back to the gas phase. The CO reacts with the N$_2$H$^+$ and forms HCO$^+$ \citep{2004LeeEvolution}. Therefore, that the N$_2$H$^+$ in our observations peaks at the continuum emission supports the proposed very young evolutionary state of this source.\\

In the early stages of the collapse either in the FHSC stage or right after protostar formation, we expect infall motions to dominate the envelope kinematics up to at least 5000 AU \citep{2013BellocheObservation}. Rotation can be expected as well but with rotation velocities being much smaller than infall velocities. At later times, we expect the outflow to interact with the inner envelope, adding more complexity to the velocity field and producing regions of enlarged linewidths. Several of the Class 0 sources from \cite{2007ChenOvro}  and \cite{2011TobinComplex} show evidence suggestive of outflow impact in the \nthp\ central velocities and/or linewidths maps. For instance, they show small regions with broad linewidths located at a distance from the protostar, and with a position angle consistent with the outflow lobes direction. The L1451-mm central velocity maps of \nthp\ shows ordered rather than complex central velocity maps. There is a visually clear velocity gradient perpendicular to the outflow and a ``blue bulge'' indicative of infall motions close to the protostar (Figure~\ref{fig:l1451-mmvel}). The linewidth map of this molecule shows one broad region at the position of the protostar (Figure~\ref{fig:l1451-mmsigma}). This broad linewidth region is due to the double peak profile produced by infall motions. Thus, we do not see outflow interaction in the \nthp\ observations at 1000 AU scales. 

Our observations show that the \nhtd\ could be tracing outflow motions since the linewidth map shows a peak spatially consistent with the direction of the blue outflow lobe. Broad NH$_3$(1,1) linewidths regions produced by outflow/envelope interactions have been observed in the Class 0 protostar HH211 \citep{2011TannerDynamics}. However, L1451-mm outflow has not been fully resolved at these scales, which makes difficult a clear comparison between the outflow lobe position and the distribution of linewidths seen in Figure~\ref{fig:l1451-mmsigma}. In addition, the entire region of increased linewidths seems to be much better aligned with the semi-minor axis of the envelope rather than the outflow axis, suggesting the increase in linewidths is due to increased infall/rotation motions towards the center of the envelope. 

All the carbon-bearing molecules in our sample show an increase of non-thermal linewidth towards the center. Although we can not rule out that outflow motions are producing these linewidths, it is likely that they are a product of infall rather than outflow motions. The reasons are that the \hcn\ spectrum shows an inverse p-cygni profile, indicative of infall, at the continuum position and that for \hcn, \hcop, and \cs\ the blueshifted emission is more intense, compact, and centered than the redshifted emission (Figure~\ref{fig:l1451-mmvel}), which would be the product of blue asymmetric lines. 

The fact that we do not see any sign of an outflow (or its impact on the envelope) it is evidence against the presence of a larger outflow component not seen by \citealt{2011PinedaEnigmatic} observations. Even if we were to assume that the increased linewidth in the \nhtd\ and the carbon-bearing molecules towards the center of the maps in Figure~\ref{fig:l1451-mmsigma} are due to an extended component of the outflow missed by \cite{2011PinedaEnigmatic}, it would imply that the outflow would extend to a maximum of 10" or 2300 AU.  At the observed outflow characteristic velocity this would result in a dynamical age of 8400 years (instead of the 2,000 years estimated in \citealt{2011PinedaEnigmatic}), which is still within range expected for the age of a first core or a very young Class 0 protostar. Furthermore, deriving a dynamical age this way assumes that all gas is launched from the protostar.  If most of the gas is entrained locally by a faster underlying wind, these dynamical ages are overestimates of the real timescales.

\subsection{Mass Infall Rate versus Accretion Rate}

We used a toy model to probe the envelope kinematics using the \nhtd\ position velocity map at a cut perpendicular to the outflow direction. We found that the emission could be well reproduced by our toy model in which infall motions dominate over rotational motions, and angular momentum is conserved. We found that the best agreement to the emission was obtained for a central mass of 0.06 M$_{\odot}$. This value could be smaller due to the self-gravity of the envelope, and/or could be up to a factor of a few larger due to the slowing down of the infalling material in the presence of magnetic fields. Nevertheless, the estimated mass for the central object is very small and consistent with either a very young protostar or a first core. From the fitting of the infall profiles in the \nthp\ and \hcn\ spectra at the continuum position we estimate an infall velocity of $\sim0.17$ \kms. Using this velocity and the LTE mass of the envelope ($\sim$ 0.2-0.3 M$_{\odot}$) within about one beam radius ($\sim$1600 AU) centered at the continuum peak\footnote{This is the approximate size of the region where the infall profile is detected (see Figure~\ref{fig:L1451-mmhole}).} position, we estimate a mass infall rate of $\dot{M}_{inf} \sim$ M$v/r$ $\sim5-7\times10^{-6}$ M$_{\odot}$ yr$^{-1}$, comparable to values found in low luminosity Class 0 protostars (\citealt{1997OhashiInterferometric,2002BellochMolecular}). \cite{2011PinedaEnigmatic} modeled the SED and visibilities for L1451-mm assuming that the central source was a protostar with a disk and derived a mass infall rate of $7.1\times10^{-6}$ M$_{\odot}$ yr$^{-1}$ which is in a good agreement with our estimate.  Using the estimated mass of the central object $\sim0.06$ M$_{\odot}$ and the inferred mass infall rate we obtain an age of $\sim10^{4}$ years for the central source, consistent with a very young protostar or a first core. More interestingly, we can use $L_{int}\leq 0.016$ L$_{\odot}$ \citep{2011PinedaEnigmatic} as an upper limit to the accretion luminosity which gives us a lower limit on the size of the central object using the following equation:
\begin{equation}
\label{eq:size}
\frac{GM\dot{M}_{acc}}{R}=L_{acc}\leq L_{int}\leq 0.016 L_{\odot}   
\end{equation}
For this estimation, we assume that the accretion rate onto the central object $\dot{M}_{acc}$ is the same as the mass infall rate $\dot{M}_{inf}$ inferred above. The lower limit for the size of the central core is then estimated to be 1.5-5 AU, which is too large to be a protostar and consistent with simulations of first cores. Any uncertainty in the central mass, either due to self-gravity or magnetic fields, does not contribute enough to change the fact that the mass infall rate inferred from the observations is too large to explain the low internal luminosity, except in the first core or episodic accretion scenarios (see below). If we were to assume a typical radius for a protostellar object ($\sim3$ R$_{\odot}$) for the size of the central object, we would then need $\dot{M}_{acc}$ to be $\lesssim2-4\times10^{-8}$ M$_{\odot}$ yr$^{-1}$, in order for $L_{acc}$ to be $\lesssim 0.016$ L$_{\odot}$ (using Eq.~\ref{eq:size}). In this case $\dot{M}_{acc} \ll \dot{M}_{inf}$, which could be explained if gas from the envelope is being accumulated in a disk around the protostar. This is what is expected in the case of episodic accretion; infall rates from the envelope onto the disk are much higher than the accretion rate from the disk onto the star, material in the disk accumulates and high accretion burst are produced by gravitational instabilities in the disk. \cite{2012DunhamResolving} studied episodic accretion using simulations and in those models, the accretion variability starts a few $10^{4}$ yrs after protostar formation, which is when the disk develops instabilities. Periods of accretion below 10$^{-7}$ M$_{\odot}$ yr$^{-1}$ occur only after several 10$^{4}$ yrs, when the protostar has accumulated at least 0.1 M$_{\odot}$. The only scenario in the simulations presented by \cite{2012DunhamResolving} in which a protostar of 0.06 M$_{\odot}$ could be accreting at such a low rate (and with a high infall rate onto the disk) is a scenario where the disk formed very recently (in last few 10$^{3}$ yrs) and it just started to accumulate mass. In these stages the disk would have a mass less than a few 0.01 M$_{\odot}$. \cite{2015TobinSub} analyzed 1.3mm interferometric continuum observations and found that at 0.3" resolution the emission is still unresolved. However, their data is consistent with there being a compact component of about 0.011 M$_{\odot}$. One possible interpretation is that this compact component is partly due to a disk. Thus, an extremely young protostar with a recently formed disk remains a possible explanation for the low luminosity of this source. The above simulations though, used a sink particle with a size of 6 AU, which is consistent with the expected size of a first core, and do not really resolve the accretion onto the protostar. 

The simulations in \citealt{2015TomidaRadiation} (see also \citealt{2011MachidaOrigin}), which have the resolution to resolve the first core, show that depending on the level of initial rotation it is possible that the first core becomes the circumstellar disk before or a few years after the protostar is born. Thus, the early disk in these cases is more massive than the protostar and can show instabilities \citep{2015TomidaRadiation}. The simulations predict that the mass of the first core/circumstellar disk is in the range of 0.02-0.08 M$_{\odot}$ with a size of 1-9 AU, and the protostellar mass (if born) is only a few $10^{-3}$ M$_{\odot}$ at that stage, around 3 years after protostar formation. The predicted mass and size for the circumstellar disk/first core are consistent with our estimate for the central source in L1451-mm. 

\subsection{Outflow evidence}

Previous observations in CO(2-1) detected a very compact ($\lesssim500$ AU) and low velocity ($2$ \kms) outflow launched by L1451-mm \citep{2011PinedaEnigmatic}. Our molecular line maps show no indication of a more extended outflow component, missed in these previous observations. Likewise, no infrared emission related to a high-velocity wind have been detected (see deep IR images in \citealt{2006FosterCloudshine}). These observations indicate that the outflow from this source is indeed very young and additionally, consistent with simulations of first cores (\citealt{2008MachidaFormation,2013TomidaRadiation,2014BateCollapse,2015TomidaRadiation}). For instance, \citealt{2008MachidaHigh} showed that the velocity of the wind launched by a central object can be estimated as the Kepler velocity at the surface of the structure from where the flow is being launched. With the estimated mass of $\sim0.06$ M$_{\odot}$ we obtain a wind velocity of 4-5 \kms\ for a first core of 1.5-5 AU and 50-70 \kms\ for a protostar of 3 R$_{\odot}$. The later case is inconsistent with the present observations, considering that the dynamical time of the observed outflow ($2000$ years) is large enough for a wind of a few tens of \kms\ to have crossed the outer regions of the core. Another scenario that could explain these observations is one in which there is a protostar with an outflow being lunched from its surrounding disk a few AU from the protostar.

Thus, the observation that the outflow from L1451-mm seems to be indeed compact and low in velocity is evidence for the central source being very young, but is not sufficient evidence for L1451-mm being a first core, specially since there are known young protostars with similar outflow properties (e.g. \citealt{2006BourkeSpitzer,2013TakahashiDirect}). However, unlike L1451-mm, the SED from these other young sources reveal a more evolved stage, consistent with being protostellar.

\section{Summary and conclusions}

In order to constrain the evolutionary state of the first hydrostatic core candidate L1451-mm, we studied the kinematic properties of its envelope at 1000 AU scales. We used CARMA observations of molecular emission lines (NH$_2$D, N$_2$H$^+$, HCN, HCO$^+$ and CS) and continuum at 3mm. Our results and conclusions can be summarized as follows.

\begin{enumerate}

\item
As compared with the samples of Class 0 sources observed in \nthp\ at a similar resolution (\citealt{2007ChenOvro,2011TobinComplex}), L1451-mm seems younger based on the observation that (a) the \nthp\ shows no sign of destruction by CO released back to the gas phase after protostar formation and (b) the broad linewidths regions are consistent with being produced by infall/rotation motions, tentative outflow/envelope interactions are seen only at scales $<2300$ AU. 

\item
\hcn\ and \nthp\ show clear signature of infall motion in their spectrum at the position of the continuum source. The \hcn\ shows an inverse p-cygni profile and the \nthp\ shows a double peak profile with blue asymmetry. We fit both spectra using the HILL model with the addition of a continuum source. We inferred an infall velocity of $\sim$ 0.17 \kms. This value for the infall velocity agrees with values observed in Class 0 sources and also with simulations of first cores at 1000 AU scales. We also find evidence of infall motions as blueward velocity shifts in the central velocity map of \nhtd, \nthp, \hcn, \hcop\ and \cs.

\item
The \nhtd\ and \nthp\ central velocity maps show gradients along a direction that is approximately perpendicular to the outflow direction. We modeled the position velocity map of the \nhtd, at a cut perpendicular to the outflow direction. The model considered simple ballistic kinematics, in which infall dominated over rotation and angular momentum is conserved. The observations show good agreement with our model, and from this we inferred a mass of $\sim0.06$ M$_{\odot}$ for the central source. We caution that this estimate could be uncertain (at most) by a factor of a few due to the simplicity of our model.

\item
Using the LTE mass from the \nthp\ emission plus the infall velocity within 1610 AU of the continuum source we inferred a mass infall rate $\dot{M}_{inf}$ of about $6\times10^{-6}$ M$_{\odot}$ yr$^{-1}$, comparable with low luminosity Class 0 protostars in the literature. From the inferred mass and the mass infall rate we estimate an age of $10^{4}$ yrs for this source. This value is close to the maximum lifetime of a first core predicted in simulations (\citealt{2010TomidaExposed}). Assuming that L$_{acc}\lesssim$ L$_{int}$ and that $ \dot{M}_{inf}=\dot{M}_{acc}$ we inferred a lower limit for the size of the central source of $\sim$1.5-5 AU, consistent with simulations of first cores. In order for the source to have a size that is consistent with that of a protostar ($\sim3$ R$_{\odot}$) and have the observed luminosity it would need to be accreting with a rate 100 times smaller than the observed mass infall rate at 1000 AU, regardless of our uncertainty in the estimated central mass. The low accretion onto the protostar could be explained if material is being accumulated in a recently formed disk surrounding a very young protostar.

\item
According to simulations, if we assume our estimated mass for the central object,  the velocity and the observed extent of the outflow are consistent with a launching radius of 1.5-5 AU (i.e., first core scenario). We note, however, that evidence from the outflow alone is not enough to distinguish if it is a first core or a protostar.
\end{enumerate}

The results of this kinematic study lead to the conclusion that L1451-mm is indeed in a very early stage of evolution, either a first core or an extremely young Class 0 protostar. Given the estimated mass, age and size of the central object, the small size of the outflow and the non-detection at mid-IR wavelengths with Spitzer \citep{2011PinedaEnigmatic}, L1451-mm is the most promising first core candidate out of the current list of FHSC candidates. Further observations at a higher resolution are needed in order to properly resolve the compact outflow and study if there is a disk in the inner 100 AU. These observations would be useful for setting strong constraints on the true evolutionary state of L1451-mm as well as observational evidence regarding the early formation and evolution of the protostar/disk/outflow system.

\acknowledgments
The authors thank the anonymous referee for a careful reading and a number of useful comments which improved the paper as a whole. MJM thanks the supports from CONICYT PAI/INDUSTRIA 79090016, XC acknowledges support by the NSFC through grant 11473069 and JEP acknowledges the financial support of the European Research Council (ERC; project PALs 320620).

\bibliography{firstcorelibrary}

\end{document}